\documentclass[aps,pre,twocolumn,amsmath,amssymb,showpacs,amsfonts]{revtex4}
\usepackage{epsfig}
\usepackage{graphicx}
\usepackage{dcolumn}
\usepackage{bm}
\usepackage{color}
\usepackage{epstopdf}

\begin{document}
\title{Solvable continuous time random walk model of the motion of tracer particles through porous media}
\author{Itzhak Fouxon$^{1,2}$}
\author{Markus Holzner$^1$}
\affiliation{$^1$ Institute of Environmental Engineering, ETH Zurich, Wolfgang-Pauli-Strasse 15, 8093 Zurich, Switzerland}
\affiliation{$^2$ Institute of Mechanical Science, Vilnius Gediminas Technical University, J. Basanavičiaus str., 28, 03224 Vilnius, Lithuania}

\begin{abstract}

We consider the continuous time random walk model (CTRW) of tracer's motion in porous medium flows based on the experimentally determined distributions of pore velocity and pore size reported in \emph{Holzner et al. Phys. Rev. E 92, 013015 (2015)}. The particle's passing through one channel is modelled as one step of the walk. The step's (channel) length is random and the walker's velocity at consecutive steps of the walk is conserved with finite probability mimicking that at the turning point there could be no abrupt change of velocity. We provide the Laplace transform of the characteristic function of the walker's position and reductions for different cases of independence of the CTRW's step's duration $\tau$, length $l$ and velocity $v$. We solve our model with independent $l$ and $v$. The model incorporates different forms of the tail of the probability density of small velocities that vary with the model parameter $\alpha$. Depending on that parameter all types of anomalous diffusion can  hold, from super- to subdiffusion. In a finite interval of $\alpha$, ballistic behavior with logarithmic corrections holds that was observed in a previously introduced CTRW model with independent $l$ and $\tau$. Universality of tracer's diffusion in the porous medium is considered.  \end{abstract}

\maketitle
\section{Introduction}

Based on the seminal work of Montroll and Weiss~\cite{MontrollWeiss} on the theory of random walks on lattices, Scher and Lax~\cite{Scher} extended the CTRW theory and incorporated memory effects to solve impurity conduction in disordered solids. Since then, the CTRW theory has been applied to model a broad range of problems characterized by history-dependent dynamics in fluctuating and disordered systems \cite{Klafter,Metzler2001}. In porous media, the CTRW model has been successfully used to predict anomalous transport \cite{Berkowitz1995,Berkowitz1997,Berkowitz2006,Metzler2001,Dentz2004,Dentz2010,Lester2013,Lester,Margolin2002}. For uncorrelated disorder, space and time increments are uncoupled and the resulting anomalous transport has been considered via decoupled CTRW's  \cite{Berkowitz2006,Metzler2001,Dentz2004,Margolin2002}. Spatial correlation gives rise to coupling which has been modeled using fully coupled CTRW's \cite{Montero2007,Dentz2010} and correlated CTRW's \cite{LeBorgne2008,Tejedor2010,Kang2011,DeAnna2013}. Correlations between successive waiting times were shown to give rise to subdiffusion even when they are Gaussian, while correlations between jump lengths produced superdiffusion \cite{Tejedor2010}. Recent work showed that correlations in the Lagrangian velocity (in particular slow velocities) produced superdiffusion \cite{LeBorgne2008,Kang2011,DeAnna2013}.

Anomalous transport arises from heterogeneous pore scale velocity which in turn depends on the pore space geometry~\cite{Berkowitz2006,bouchaud,Bijeljic,DeAnna2013,Markus,Kang2014,Siena2014}. In several papers, pore scale velocity PDFs in porous media have been found to have positive (exponential or stretched exponential) tails \cite{Bijeljic,Cushman2001,Datta2013,DeAnna2013,Markus,Kleinfelter2005,Kutsovsky1996,Lachhab2008,Lebon1996,Moroni2001,Moroni2001B,Neel2014,Seymour2004,Siena2014}. In \cite{Markus}, the heavy-tailed velocity PDF could be related to the (exponential) pore-size distribution and flow connectivity. A key element was the introduction of a CTRW model that could explain the observed non-Fickian transport behaviors~\cite{Markus}. However, despite significant progress, the relation between pore scale geometry, intermittent pore-scale flow and non-Fickian transport remains not fully understood. In this work we consider the CTRW model based on a given distribution of step lengths and provide exact solution for the resulting transport behavior of flow particles.

\subsection{Pore-scale CTRW model}

In general, CTRW models are effective models characterized by distributions of transition times, lengths and velocities. At the pore scale, the relevant distributions are pore length and pore velocity, while for a fracture network one has to consider the distribution of fracture lengths etc. Ref.~\cite{Bijeljic2006} considered for the first time a pore network where a CTRW is parameterized using the distribution of pore velocities. The CTRW model studied in this work is slightly simplified version of the model introduced in \cite{Markus} based on direct modelling of experimental observations. A fluid tracer in a real porous medium like soil finds itself either inside a well-defined unique channel or at a junction between the channels. For the latter it may be difficult to find a unique channel to which the tracer can be assigned (the number of the channels connected to the given junction is determined by the connectivity of the junction). The model assumes that this assignment is done in some way so that the tracer is always found in a certain definite pore. This does not imply that junctions are neglected. In fact stagnation regions present at some junctions cause long stays of the tracer. Those stays could cause anomalous transport at large times so the considered property cannot be disregarded in studies of anomalous transport. The model includes these long stays in the definition of residence times inside the channel to which the tracer is assigned.

Thus the motion consists of a sequence of finite time-intervals so that during a given interval the tracer belongs to a certain fixed channel. The model considers the passage from channel to channel as instantaneous event. Motion inside the channel is determined by the random length of the channel and the time spent inside the channel which is the length divided by the random velocity. This velocity is considered as positive quantity whose temporal variation inside the channel is neglected. The model does not study the directions of motion as determined by the pores' orientation in space but only lengths of the channels and times spent in them. This does not signify considering the medium as a sequence of singly connected channels. Rather the model studies the absolute distance passed by the particle but not the net displacement in space (the relation between the two is determined by tortuosity discussed in the final Section). The way the channels are organized in space is outside the domain of consideration of the model.

The core experimental observation of \cite{Markus} is that when the particle reaches the end of the channel it can either smoothly pass to the next channel without changing the velocity strongly or undergo strong acceleration. The value of this acceleration is determined by the complex geometry of the pores and is to be considered random. It is this way of motion that is proposed to be the key property of motion in the porous medium that determines the anomalous diffusion.

We propose that the model where the particle keeps its velocity constant between instantaneous random acceleration events produces the same laws of growth of moments of the distance as the real medium. For instance in the model the dispersion of the passed distance is found below to obey a power law in time. It is proposed then that in the porous medium the growth will obey the power-law with the same exponent but different prefactor.

It was observed in \cite{Markus} that acceleration events occur typically once per ten passages between the pores. Thus at a typical passage the magnitude of the particle's velocity does not change much (the direction's change is not relevant for the questions considered in this paper). The passage through one pore was modelled as one step of the walk. The length of this step is the random length of the pore. The velocity at the next step can be conserved with probability $\lambda$ (if there is no acceleration event) or refreshed with probability $1-\lambda$ (if there is an acceleration event) with the new value taken from the experimentally motivated  probability density function (PDF) of velocities \cite{Markus}. Such a persistence parameter $\lambda$ was also used in \cite{Kang2015,Kang2015b}. This persistence implied an effective longer transition length that is re-scaled by $1/\lambda$ \cite{Markus}. The velocity of motion inside the pore is the half sum of the velocities at the pore's ends. In our CTRW the length of one step does not obey a broad distribution, cf. \cite{bouchaud}, but rather has an exponential distribution. The anomalous diffusion arises because one step can take very long time. This is because of near trapping of particles in some pores. That is reflected via the power-law tail of the PDF of the step's duration, see below.

It was demonstrated that this model provides realistic description of the observations. Here, we introduce a simplification taking the velocity inside the pore not as half sum but as the velocity at the pore's entrance. This simplification does not seem to be of consequence for the laws of anomalous diffusion that concern us here.

\subsection{Separation of variables}

Due to the mentioned persistence of velocity, the described model has correlations between the walk's steps which makes it hard for theoretical treatment. However, we observe that redefining the steps the model can be reduced to CTRW with independent consecutive steps. We redefine the step as the motion between consecutive acceleration events. Thus by definition velocities and lengths at the consecutive steps are independent. The PDF of the step's length undergoes the corresponding "renormalization" where the step's length is sum of the lengths of the pores passed without changing the velocity. Thus the more usual setting with independent consecutive steps is recovered.

The CTRW that we find is separable: the step length $l$ and the velocity $v$ during the step are independent random variables. For a general CTRW the step is characterized by three variables: the duration of the step $\tau$, the constant velocity $v$ at which the step is performed and the spatial displacement $l$ during the step (where in dimension higher than one $\bm v$ and $\bm l$ are vectors). We have $l=v\tau$ so only two of the three variables are independent. We call the CTRW separable if any pair of the step's variables are independent, cf. \cite{bouchaud}. Thus there are three types of separable CTRWs. The walk with independent $\tau$ and $v$ is called L\'{e}vy walk, see e. g. \cite{zdk}. We call the walk with independent $l$ and $v$ the $l-v$ CTRW and the walk with independent $l$ and $\tau$ the $l-\tau$ CTRW. In \cite{bouchaud} the $l-\tau$ CTRW is called separable CTRW, our use of the term "separable" is different.

The three described separable walks are quite different. In our separable $l-v$ CTRW model of the tracer's motion in the porous medium the distribution $p(l)$ of length $l$ is fast decaying. The fast decay of $p(l)$ seems to be the necessary property of the porous medium (disregarding the possible existence of long sequences of almost parallel channels - corridors - over which the tracer's velocity does not change much. The study of those is beyond our scope here). Thus the fluctuations of $l$ are weak so that fluctuations of step's duration $\tau=l/v$ are chiefly those of $1/v$ (which is independent of $l$). Correspondingly if the probability density function (PDF) of velocity is finite at zero velocity then $\langle \tau\rangle\propto \langle 1/v\rangle$ diverges. This implies that the PDF of $\tau$ has a power-law tail with decay exponent smaller or equal $2$. This is the known reason for anomalous diffusion defined as power-law growth of the coordinate's dispersion whose exponent differs from one \cite{bouchaud}. The laws of anomalous diffusion that we find seem to be richer than for $l-\tau$ CTRW or L\'{e}vy walk, see \cite{bouchaud} and \cite{inf} respectively.

In L\'{e}vy walk anomalous diffusion is found in the case where the PDF of $\tau$ has a power-law tail and the PDF of velocity is fast decaying. When the tail's decay exponent is between $1$ and $2$ so $\langle \tau\rangle=\infty$ the dispersion of the particle's position has universal quadratic growth in time. For exponent between $2$ and $3$ one finds superdiffusion where the dispersion grows faster than linearly in time but slower than quadratically in time. Finally for exponent larger than $3$ normal diffusion holds where dispersion grows linearly in time. Thus in this case only superdiffusion holds \cite{inf}.

In contrast in $l-\tau$ CTRW when the PDF of $\tau$ has a power-law tail with exponent between $1$ and $2$ the dispersion grows slower than linearly in time (subdiffusion). For an exponent between $2$ and $3$ normal diffusion with linear growth of dispersion holds in the leading order. Thus in this case only subdiffusion holds \cite{bouchaud}.

 Our $l-v$ CTRW can give both super- and subdiffusive growths of the dispersion, as previously noted in \cite{Margolin2002,Dentz2004}. These are determined by the behavior of the PDF of velocity $p_v(v)$ at small velocities. For finite $p_v(0)$ where $\langle\tau\rangle=\infty$ we find ballistic growth with logarithmic correction. Up to that correction this is similar to the L\'{e}vy walk in the regime of divergent $\langle \tau\rangle$. However when $p_v(v)$ has integrable power-law singularity at zero velocity we find that arbitrary growth exponent between $0$ and $2$ is possible depending on the power-law's exponent. Thus this $l-v$ CTRW can incorporate both superdiffusion and subdiffusion.

In the case of finite $p_v(v)$ that leads to $\tau^{-2}$ tail of the PDF of $\tau$ we find that the average distance passed by the particle grows as $t/\ln t$ and the dispersion of the distance as $t^2/ln^3 t$. Though these laws hold for the distance (path length) rather than spatial displacement when the geometry of the medium is not too complex these laws can be transferred from distance to the displacement of the particle in the direction of the mean flow (in simplest case this is done introducing projection factor in the direction of the flow).

Identical $t/\ln t$ and $t^2/\ln^3 t$ laws of growth were obtained in $l-\tau$ CTRW model of the tracer's motion in the porous medium introduced in \cite{Lester}. The studied PDF of the step's duration had $\tau^{-2}$ tail identical with that described above, see the discussion in the final section. We hence note that different types of separable CTRW can produce either similar or different results.

\subsection{Outline}

In the next Section we derive for inseparable CTRW the Laplace transform of the characteristic function of walker's coordinate through statistical properties of one step of the walk. The corresponding formula is known as the Montroll-Weiss equation (this equation itself will not be needed until Section \ref{dif}). Considering the reductions of the equation for different types of separable walks we clarify the difference between different types of separations. Though the results are scattered over the literature it seems that they were not concentrated in one place and the difference between different types of separability was not stressed. This difference is to be considered carefully in modelling transport in the porous medium.

Having provided the relevant basis for the study of CTRWs we describe our model of tracer motion in the porous medium in Section \ref{model}. The model was introduced phenomenologically based on the experiment. This model is not a CTRW because the velocity in subsequent pores is correlated with finite probability. However in the next Section we demonstrate that the model can be reduced to a CTRW by redefining the step. We find that we can introduce an effective length of the pore beyond which the velocity decorrelates. This length is about ten times the typical length of one pore. We expect this to be realistic feature of the porous medium as discussed in the final section.

The key property that underlies the anomalous diffusion at large times is the power-law tail of the PDF of the residence time in the pore derived in Section \ref{power}. The power-law tail tells that the tracer can spend much time in the pore with quite high probability. In the real porous medium long residence times can be caused by stagnation regions at the junctions and/or long stays near the channels' walls where velocity of the fluid is very small because of the no-slip boundary condition at the walls. These delays in the propagation through the medium cause anomalous scaling of passed distance with time.

Besides the passed distance, porous medium flow can be characterized by a different random variable which is the number of pores passed by the particle in time $t$. We find this quantity very useful because it is simpler for calculations than the distance but many properties of the distance statistics can be inferred from it. The reason is that the length of the single pore does not have strong fluctuations so the passed distance for many purposes is equivalent to the number of passed pores times a characteristic pore length. The calculation of the PDF of the number of passed pores is done in Section \ref{pd}. We find the power-law growth of the average number of passed pores and the dispersion of that number with time. In the next Section we demonstrate that this power-law growth coincides with the power-law growth of the passed distance where the difference between the two random variables is in the prefactor of the power. The derivation is based on using the Montroll-Weiss equation for the CTRW formulation of our model. Finally, in the final discussion section we provide the formulation of the results and their implications.

\section{Inseparable CTRW's propagator and reductions for separable walks} \label{insep}

In this Section we describe the framework of CTRW and provide the Fourier-Laplace transform of the PDF $p(t, \bm x)$ of the position $\bm x(t)$ of the particle ("propagator"). The CTRW is determined by the law of change of $\bm x(t)$,
\begin{eqnarray}
&& \!\!\!\!\!\!\!\!\!\!\!\! \bm x_{n+1}=\bm x_n+\bm l_n,\ \ t_{n+1}=t_n+\tau_n,\ \ \bm l_n=\bm v_n\tau_n,
\end{eqnarray}
where $\bm x_0=0$, $t_0=0$ and between the steps the particle moves at constant velocity $\bm v_n$. The quantities pertaining to different steps of the walk - $\bm l_n$, $\tau_n$, $\bm v_n$ - are considered independent.

The inseparable CTRW is determined by the joint PDF of any pair of the random variables $\bm l$, $\tau$, $\bm v$ characterizing one-step statistics. We take the pair to be $\tau$ and $\bm l$ and designate the corresponding PDF by $\psi(\tau, \bm l)$ using the notation similar to that of \cite{bouchaud,Berkowitz2006}.

The derivation of $p(t, \bm x)$ is done introducing auxiliary probability density $q(t, \bm x)$ that the particle finishes one of the walk's steps in the vicinity of $\bm x$ at time $t$. thus $q(t, \bm x)d\bm xdt$ is the probability that one of the walk's steps finishes in time interval $(t, t+dt)$ in $d\bm x$ vicinity of $\bm x$. We observe that the particle's position at the beginning of the step that ends at $\bm x$ at time $t$ can be either $\bm x=0$ at time $t=0$ or a certain position $\bm x'$ at time $t-\tau$ where $0<\tau<t$. In the first case the particle comes to $\bm x$ in one step directly from its initial position. This contributes $\psi(\tau, \bm x)$ to $q(t, \bm x)$. In the case that the particle performed more steps before ending up at $\bm x$ the corresponding contribution to $q(t, \bm x)$ is the product of probability of reaching $\bm x'$ at time $t-\tau$ and the probability of making a step of duration $\tau$ from $\bm x'$ to $\bm x$. Summing the probabilities we find,
\begin{eqnarray}
&& \!\!\!\!\!\!\!\!\!\!\!\!\!\! q(t, \bm x)\!=\!\psi(t, \bm x)\!+\!\int_0^t \!\!d\tau\int \!\!d\bm x' q(t\!-\!\tau, \bm x') \psi(\tau, \bm x\!-\!\bm x').\label{qt}
\end{eqnarray}
If we introduce $q(t, \bm x)=\nu(t, \bm x)-\delta(t)\delta(\bm x)$ then we find,
\begin{eqnarray}
&& \!\!\!\!\!\!\!\!\!\!\!\!\!\! \nu(t, \bm x)\!=\!\!\int_0^t \!\!d\tau\int \!\!d\bm x' \nu(t\!-\!\tau, \bm x') \psi(\tau, \bm x\!-\!\bm x')+\delta(t)\delta(\bm x). 
\end{eqnarray}
(Here the $\delta-$function is defined so that $\int_0^{\infty}\delta(t)dt=1$). The functions $q(t, \bm x)$ and $\nu(t, \bm x)$ coincide at $t>0$ and any of them can be used as intermediate quantity in the derivation of the Montroll-Weiss equation. We use the function $q(t, \bm x)$ that provides the continuum counterpart of the corresponding equation on the lattice, see e. g. \cite{bouchaud}. The consideration using $\nu(t, \bm x)$ and the corresponding equation on the PDF is used e.g. in \cite{Metzler2001}.

The solution of Eq.~(\ref{qt}) is found using the Fourier-Laplace transform in coordinate and time respectively using
\begin{eqnarray}
&& \!\!\!\!\!\!\!\!\!\!\!\! \psi(s, \bm k)=\int_0^{\infty}dt\int d\bm x  \exp[-st-i\bm k\cdot\bm x]\psi(t, \bm x),
\end{eqnarray}
with similar formulas for other functions of $t$ and $\bm x$. We designate the function and its transform by the same letter so the distinction is done via the argument. We find,
\begin{eqnarray}
&& \!\!\!\!\!\!\!\!\!\!\!\! q(s, \bm k)=\frac{\psi(s, \bm k)}{1- \psi(s, \bm k)}.\label{q}
\end{eqnarray}
The probability $p(t, \bm x)d\bm x$ of finding the particle at time $t$ in $d\bm x$ vicinity of $\bm x$ is the sum of the probabilities of reaching that volume
before and after the end of the first step of the walk. The probability of passing $d\bm x$ vicinity of $\bm x$ during the first step is $f(t, \bm x) d\bm x$
where we defined,
\begin{eqnarray}&& \!\!\!\!\!\!\!\!\!\!\!\! f(t, \bm x)=\int_t^{\infty} \left(\frac{\tau}{t}\right)^d \psi\left(\tau, \frac{\tau \bm x}{t}\right)d\tau .
\end{eqnarray}
with $d$ the dimension of space. Here we observed that the particle that finished the first step of the walk at time $\tau>t$ in $(\tau/t)^d d\bm x$ vicinity of $\tau \bm x/t$ was at time $t$ in $ d\bm x$ vicinity of $\bm x$. The integral over $\tau$ includes all possible times of finishing the first step of the walk. We find for the PDF using similar consideration for coming to $\bm x$ after finishing the last step of the walk at $\bm x'$ that,
\begin{eqnarray}&& \!\!\!\!\!\!\!\!\!\!\!\! 
p(t, \bm x)\!=\!f(t, \bm x)\!+\!\int_0^t\!\! d\tau \int\!\! d\bm x' q(t\!-\!\tau, \bm x')f(\tau, \bm x-\bm x'),\label{basicp}
\end{eqnarray}
Fourier-Laplace transform using Eq.~(\ref{q}) gives,
\begin{eqnarray}
&& \!\!\!\!\!\!\!\!\!\!\!\! p(s, \bm k)=\frac{f(s, \bm k)}{1-\psi(s, \bm k)}.\label{psk}
\end{eqnarray}
Thus we consider $f(s, \bm k)$ that obeys,
\begin{eqnarray}
&& \!\!\!\!\!\!\!\!\!\!\!\! f(s, \bm k)\!=\!\int_0^{\infty}\!d\tau \int_0^{\tau}\! dt\int \exp\left[-st-\frac{it\bm k\cdot\bm l}{\tau}\right] \!\!\psi\left(\tau, \bm l\right)d\bm l, 
\end{eqnarray}
which integration gives,
\begin{eqnarray}
&& \!\!\!\!\!\!\!\!\!\!\!\!\!\!\!\!\! f(s, \bm k)\!=\!\int_0^{\infty}\!d\tau\int\!  d\bm l
\frac{\psi\left(\tau, \bm l\right)\left(1-\exp\left[-s\tau-i\bm k\cdot\bm l\right]\right)}{s+i\bm k\cdot\bm l/\tau}.\!\! 
\end{eqnarray}
This together with Eq.~(\ref{psk}) gives the solution for the Fourier-Laplace transform of the characteristic function of the position of the inseparable CTRW's walker. We provide below another form of the solution and some reductions.

We can write $p(s, \bm k)$ using the joint PDF $\psi'(\tau, \bm v)$ of $\tau$ and $\bm v$ instead of $\psi(\tau, \bm l)$ which gives a somewhat nicer form. Introducing the integration variable of velocity $\bm v=\bm l/\tau$ and using $\psi'(\tau, \bm v)=\tau^d \psi(\tau, \bm v\tau)$ we write,
\begin{eqnarray}
&& \!\!\!\!\!\!\!\!\!\!\!\! f(s, \bm k)\!=\!\int_0^{\infty}\!\!d\tau\int\!\!  d\bm v
\frac{\psi'\left(\tau, \bm v\right)\left(1\!-\!\exp\left[-\tau\left(s\!+\!i\bm k\cdot\bm v\right)\right]\right)}{s+i\bm k\cdot\bm v}\nonumber\\&&\!\!\!\!\!\!\!\!\!\!\!\! =\left\langle \frac{1}{s+i\bm k\cdot\bm v}\right\rangle_v-\int \frac{\psi'(s+i\bm k\cdot\bm v, \bm v)}{s+i\bm k\cdot\bm v}d\bm v,\end{eqnarray} where the averaging is over statistics of velocity. Similarly we have,
\begin{eqnarray}
&& \!\!\!\!\!\!\!\!\!\!\!\! \psi(s, \bm k)=\int  \psi'\left(s+i\bm k\cdot\bm v, \bm v\right)d\bm v.
\end{eqnarray}
We find using Eq.~(\ref{psk}) that,
\begin{eqnarray}
&& \!\!\!\!\!\!\!\!\!\!\!\! p(s, \bm k)=\left[\left\langle \frac{1}{s+i\bm k\cdot\bm v}\right\rangle_v-\int \frac{\psi'(s+i\bm k\cdot\bm v, \bm v)}{s+i\bm k\cdot\bm v}d\bm v\right]\nonumber\\&&\!\!\!\!\!\!\!\!\!\!\!\!\times\frac{1}{1-\int  \psi'\left(s+i\bm k\cdot\bm v, \bm v\right)d\bm v}.\label{os}
\end{eqnarray}
For L\'{e}vy walk we have $\psi'\left(\tau, \bm v\right)=\psi(\tau) p_v(\bm v)$. Here $p_v(\bm v)$ and $\psi(\tau)$ are the marginal distributions of $v$ and $\tau$ respectively,
\begin{eqnarray}
&& \!\!\!\!\!\!\!\!\!\!\!\! p_v(\bm v)=\int_0^{\infty}\psi'(\tau, \bm v)d\tau,\ \ \psi(\tau)=\int_0^{\infty}\psi'(\tau, \bm v)d\bm v.
\end{eqnarray}
where we distinguish functions by their arguments. For instance $\psi(\tau)$ is the PDF of the residence time $\tau$ but $\psi(\tau, \bm l)$ is the joint PDF of $\tau$ and $\bm l$. The use of $\psi'\left(\tau, \bm v\right)=\psi(\tau) p_v(\bm v)$ in Eq.~(\ref{os}) gives the Montroll-Weiss equation of the L\'{e}vy walk, see e.g. \cite{inf,zdk} and references therein,
\begin{eqnarray}
&& \!\!\!\!\!\!\!\!\!\!\!\! p(s, \bm k)= \left\langle\frac{1-\psi(s+i\bm k\cdot\bm v)}{s+i\bm k\cdot\bm v}\right\rangle_v \frac{1}{1-\langle \psi(s+i\bm k\cdot\bm v)\rangle_v},\nonumber
\end{eqnarray}
However this formula does not hold in $l-\tau$ CTRW with $\psi\left(\tau, \bm l\right)=\psi(\tau)p(\bm l)$ where $p(\bm l)=\int \psi\left(\tau, \bm l\right)d\tau$. We have,
\begin{eqnarray}
&& \!\!\!\!\!\!\!\!\!\!\!\! f(s, \bm k)=\int_0^{\infty}d\tau\int  d\bm l
\frac{\psi(\tau)p(\bm l)\left(1-\exp\left[-s\tau-i\bm k\cdot\bm l\right]\right)}{s+i\bm k\cdot\bm l/\tau}\nonumber\\&&\!\!\!\!\!\!\!\!\!\!\!\!\!\!=\!\int_0^{\infty}\!\!\tau^d d\tau\int \!\! d\bm v
\frac{\psi(\tau)p(\bm v \tau)\left(1\!-\!\exp\left[-\tau(s\!-\!i\bm k\cdot\bm v)\right]\right)}{s+i\bm k\cdot\bm v}, \end{eqnarray}
that does not look reducible to average over one of the variables ($\bm l$ or $\bm v$) only. Similar considerations hold for $l-v$ CTRW.

For comparison with the equation provided in \cite{bouchaud} we must consider a different definition of the walk where between the steps the walker does not move but in the end of the step instantaneously changes its position by $\bm l$ (this formulation is closer to the original formulation of the walk on the lattice and is known as jump model). In this case $q(t, \bm x)$ does not change from Eq.~(\ref{qt}) but the equation on $p(t, \bm x)$ takes a form different from that in Eq.~(\ref{basicp}) because the particle does not move between the steps. We find,
\begin{eqnarray}&& \!\!\!\!\!\!\!\!\!\!\!\! 
p(t, \bm x)\!=\!\phi(t)\delta(\bm x)+\int_0^t\!\! d\tau q(t\!-\!\tau, \bm x)\phi(\tau),
\end{eqnarray}
where $\phi(t)=\int_{t}^{\infty}d\tau \int d\bm l\psi(\tau, \bm l)$ is the probability that the walk's step lasts longer than $t$. The first term above describes the contribution of the event that the particle did not move at all during time $t$ (so $\bm x=0$ for these events to contribute $p(t, \bm x)$) and the last term describes contributions of events where the particle came to $\bm x$ at time $t-\tau$ and then stopped until time $t$. The solution is,
\begin{eqnarray}&& \!\!\!\!\!\!\!\!\!\!\!\! p(s, \bm k)\!=\!\!\frac{\phi(s)}{1-\psi(s, \bm k)}, \end{eqnarray}
which differs from Eq.~(\ref{basicp}). We observe that $\phi(s)=[1-\psi(s, k=0)]/s$ which gives,
\begin{eqnarray}&& \!\!\!\!\!\!\!\!\!\!\!\! p(s, \bm k)\!=\!\!\frac{1-\psi(s, k=0)}{s[1-\psi(s, \bm k)]}, \end{eqnarray}
reproducing the equation that can be found in \cite{bouchaud} and references therein. One can derive reductions of this equation as we did previously in the case where between the steps the particle moves at constant velocity rather than pauses the walk. In \cite{Dentz2010}, an analysis of different coupled and uncoupled velocity models with position interpolation (sometimes called `Creeper model') for different (constant and heavy-tailed) distributions of the transition length can be found.

\section{Model as direct description of the experiment} \label{model} We introduce our CTRW model of \cite{Markus} for the motion of flow particle tracers in the porous medium. The motion of the particle consists of a sequence of discrete steps where one step represents the passing through one pore. The step (pore) has random length $l$ which is drawn from the empirically observed probability density function (PDF) of pore lengths \cite{Markus},
\begin{eqnarray}&&  p(l)=\frac{4}{d}\exp\left[-\frac{4l}{d}\right], \end{eqnarray}where $d/4$ is the characteristic length of the pore that is set below to one by the choice of units of length. Particle velocities in consecutive pores are considered independent if an acceleration event occurs at the junction between the pores. Since acceleration events do not occur at all throats then with finite probability $\lambda$ the particle does not change its velocity in the passage between the channels. The observations indicate that only in about one tenth of the junctions the particles undergo strong accelerations (that is $\lambda\sim 0.9$). Further we neglect the time variation of velocity during motion in one pore. This seems to be of no consequence for the study of diffusion laws below. Namely it seems plausible that the anomalous transport exponent is independent of smooth variations of velocity inside the pore. Thus the velocity during one step of the walk is a random constant drawn from the empirically observed PDF of velocities $p_v(v)$.

Below we derive PDF of the distance passed by the particle independently of the concrete form of $p_v(v)$. For the study of the results we use the concrete form of $p_v(v)$. One of the PDFs that were proposed based on spatial averaging of Poiseille profile with given observed distribution of the maximal pore velocity is \cite{Markus},
\begin{eqnarray}
&&  p_v(v)=\frac{\Gamma\left[1-\alpha, \left(v/v_0\right)^{\alpha}\right]}{v_0\Gamma(1-\alpha+1/\alpha)}, \label{velst} \end{eqnarray}
where $v_0$ the characteristic velocity and $\Gamma(\beta, x)$ is the incomplete Gamma function, $\Gamma(\beta, x)=\int_x^{\infty} \exp[-t]t^{\beta-1}dt$. A similar velocity distribution was also considered in the solute transport model based on a CTRW approach in \cite{Comolli2016}.
The motion is assumed to be only in one direction (the positive $x-$axis below) so that $v>0$. This models the motion through the porous medium under the action of pressure gradient or gravity neglecting the more rare reversals of direction of motion \cite{Markus}. Below we pick units of time so that $v_0=1$. The moments of the distribution are $\langle v^k\rangle=\Gamma(1-\alpha+(k+1)/\alpha)/[(k+1)\Gamma(1-\alpha+1/\alpha)]$
from which the average and the dispersion can be read.
The large argument asymptotic form $\Gamma(\beta, x)\sim x^{\beta-1}\exp[-x]$ implies that the distribution's tail obeys
\begin{eqnarray}
&&\!\!\!\!\!\!\!\!\!\!\!\! p_v(v)\sim v^{-\alpha^2} \exp\left[-v^{\alpha}\right]/\Gamma(1-\alpha+1/\alpha), \ \ v\gg 1.
\end{eqnarray}
The physical demand that $p_v(v)$ decays at large $v$ as stretched exponential limits the range of $\alpha$ to $\alpha>0$. If $\alpha$ is not integer then the small argument asymptotic form of $\Gamma(\beta, x)$ implies at $v\ll 1$ that,
\begin{eqnarray}
&& \!\!\!\!\!\!\!\!\!\!\!\!p_v(v)\sim \frac{\Gamma(1-\alpha)}{\Gamma(1-\alpha+1/\alpha)}-\frac{v^{\alpha(1-\alpha)}}{(1-\alpha)\Gamma(1-\alpha+1/\alpha)}.\label{as}
\end{eqnarray}
Integrability at zero demands $\alpha(1-\alpha)>-1$ which bounds the range of possible $\alpha$ from above by $[\sqrt{5}+1]/2$. We conclude that in our model the range of physically relevant $\alpha$ is $0<\alpha<[\sqrt{5}+1]/2$.

The case of exponential tail, $\alpha=1$, (which is not described by Eq.~(\ref{as}) because $\alpha$ is integer) presents interest because it seems to agree with the experiment \cite{Markus}. In this case the incomplete $\Gamma-$function becomes the exponential integral $Ei(x)$ that is $p_v(v)=-Ei(-v)$ where $-Ei(-x)=\int_x^{\infty}\exp[-t]dt/t$. We have,
\begin{eqnarray}
&&\!\!\!\!\!\!\!\!\!\!\!\! 
p_v\sim  v^{-1} \exp\left[-v\right],\ \ v\gg 1;\ \ p_v\sim -\ln v,\ \ v\ll 1.
\end{eqnarray}
Finally the temporal change in the particle position $x$ in $n-$th step of the walk is given by,
\begin{eqnarray}
&& \!\!\!\!\!\!\!\!\!\!\!\! x_{n+1}=x_n+l_n,\ \ t_{n+1}=t_n+\tau_n,\ \ \tau_n=l_n/v_n.
\end{eqnarray}
We describe the construction of the particle's trajectory. The initial position at time $t=0$ is $x_0=0$. The motion proceeds by drawing the initial velocity $v_0$ from the distribution $p_v(v)$ and the step length $l_0$ from the exponential distribution. The residence time $\tau_0$ in this initial pore is then $\tau_0=l_0/v_0$. The length is randomly renewed at $t=\tau_0$ but the velocity can remain the same with finite probability $\lambda$ . The empirically observed $\lambda$ is $9/10$ so the probability of keeping the velocity constant is quite high. If the velocity is renewed, which happens with probability $1-\lambda$ then its new value is drawn again from $p_v(v)$. The trajectory is built by iterations of the process.

Persistence of velocity with finite probability implies that the residence times $\tau_i$ and $\tau_{i+1}$ in consecutive channels $i$ and $i+1$ are not independent. We have for the joint PDF $P(\tau_i, \tau_{i+1})$ of $\tau_i$ and $\tau_{i+1}$ that,
\begin{eqnarray}&&\!\!\!\!\!\!\!\!\!
P\!=\!\left\langle \delta\left(\frac{l_i}{v_i}\!-\!\tau_i\right)\delta\left(\frac{l_{i+1}}{v_{i+1}}\!-\!\tau_{i+1}\right)\right\rangle\!=\!(1\!-\!\lambda)\psi(\tau_i)\psi(\tau_{i+1})
\nonumber\\&&\!\!\!\!\!\!\!\!\!+\lambda \int p_v dv  \left\langle\delta\left(\frac{l_i}{v}-\tau_i\right)\delta\left(\frac{l_{i+1}}{v}-\tau_{i+1}\right)\right\rangle_{l_i, l_{i+1}},\label{indep}
\end{eqnarray}
where independent averaging in the last term describes contribution of events where  velocity is conserved over $l_i$ and $l_{i+1}$. We introduced PDF of the residence time in one pore (the definition agrees with the previous Section),
\begin{eqnarray}
&& \!\!\!\!\!\!\!\!\!\!\!\! \psi(\tau)=\left\langle \delta\left(l/v-\tau\right)\right\rangle_{l, v}=\left\langle v\delta\left(l-v \tau\right)\right\rangle_{l, v},\label{spdf}
\end{eqnarray}
where the averaging is over $l$ and $v$. The first term in Eq.~(\ref{indep}) is product of functions of $\tau_i$ and $\tau_{i+1}$ but the last one given by (we average over $l$ using $p(l)$)
\begin{eqnarray}&& \!\!\!\!\!\!\!\!\!\!\!\!
\lambda\int v^2p_v dv\exp[-(\tau_i+\tau_{i+1})v],
\end{eqnarray} is not. Thus the PDF of $\tau_i$, $\tau_{i+1}$ does not factorize and $\tau_i$, $\tau_{i+1}$ are dependent. This produces difficulties in the theoretical study of the model. Fortunately redefining the walk's steps we can reduce the model to that with independent durations of consecutive steps.

\section{Reduction to CTRW with independent steps}

We redefine the walk's step as motion between the consecutive acceleration events. That motion typically includes a finite number of channels passed at constant magnitude of the velocity. Thus velocities at different steps of the walk are independent. The PDF of the value of velocity during the step remains $p_v(v)$ but the PDF of the pore length undergoes "renormalization". The PDF $p_t(l)$ that the particle passes the total length $l$ of the pores without changing its velocity and then changes velocity is formed by the sum of contributions of events with different number of passed pores. We have,
\begin{eqnarray}
&&\!\!\!\!\!\!\!\!\!\!\!\! p_t(l)=(1-\lambda)p(l)+\lambda(1-\lambda) \int_0^{l} dl_1 p(l-l_1) p(l_1)+\lambda^2\nonumber\\&&\!\!\!\!\!\!\!\!\!\!\!\! (1-\lambda) \int_0^{l} dl_1\int_0^{l-l_1} dl_2 p(l-l_1-l_2)p(l_2) p(l_1)+\ldots.
\end{eqnarray}
We find taking Laplace transform with $s$ the Laplace transform variable that,
\begin{eqnarray}
&&\!\!\!\!\!\!\!\!\!\!\!\!\!\!\! p_t(s)=\frac{1-\lambda}{\lambda}\sum_{k=1}^{\infty}\left[\lambda p(s)\right]^k=\frac{(1-\lambda)p(s)}{1-\lambda p(s)}.
\end{eqnarray}
In the case of $p(l)=\exp[-l]$ we find $p(s)=[1+s]^{-1}$ so,
\begin{eqnarray}
&&\!\!\!\!\!\!\!\!\!\!\!\!\!\!\! p_t(s)\!=\!\frac{1-\lambda}{1+s-\lambda},\ \ p_t(l_{tot})\!=\!(1\!-\!\lambda)\exp[-(1\!-\!\lambda)l_{tot}].
\end{eqnarray} Thus persistence boils down to the increase of the effective pore length by $[1-\lambda]^{-1}$ times. In the case of $\lambda=9/10$ fitting the data of \cite{Markus} this is ten times increase so this is a significant effect.

Thus the model reduces to the CTRW with independent consecutive steps. We did not use this as original formulation of the model for keeping direct bearing with the experiment underlying the model.

Below we rescale the units of length again setting $d/[4(1-\lambda)]$ to one. Thus the PDF of step's length below is $\exp[-l]$, the distribution of velocity is $p_v(v)$ with $v_0=1$ by proper choice of units of time. For clarifying the role of parameters we restore dimensions in final formulas.

\section{Power-law tail of residence time's PDF: possible universality} \label{power}

We consider the statistics of the particle's residence time $\tau=l/v$ in one pore. The average residence time is,

\begin{eqnarray}&& \!\!\!\!\!\!\!\!\!\!\!\!
\langle \tau\rangle=\int_0^{\infty} \tau\psi(\tau)d\tau=\left\langle \frac{l}{v}\right\rangle=\langle l\rangle\left\langle \frac{1}{v}\right\rangle.\label{avrg0}
\end{eqnarray}
This diverges if $\langle v^{-1}\rangle$ does. Since divergence of $\langle v^{-1}\rangle$ would be typically the case - it holds for distributions with finite $p_v(v=0)$ and for distribution (\ref{velst}) independently of $\alpha$ - then this indicates the possible deficiency of the model discussed in the conclusion. The divergence of $\langle \tau\rangle$ indicates that $\psi(\tau)$ has a power-law tail. This is demonstrated to be true below based on observing that $\psi(\tau)$ is the derivative of the Laplace transform of velocity's PDF. The tail has universal decay exponent $2$ at $0<\alpha<1$. In the $\alpha=1$ case the $\tau^{-2}$ tail has logarithmic corrections. For $\alpha$ between $1$ and $[\sqrt{5}+1]/2$ the tail's exponent is monotonously decreasing from $2$ at $\alpha=1$ to $1$ at $\alpha=[\sqrt{5}+1]/2$. Thus we deal with CTRW where $\psi(\tau)$ has power-law tail with infinite $\langle\tau\rangle$. In L\'{e}vy walk this would imply that dispersion of the distance passed by the particles grows quadratically in time independently of $\alpha$. We will see in next Sections that for our $l-v$ CTRW this is not true. Finally we demonstrate that the Laplace transform of velocity's PDF is the probability of not leaving the initial channel during time equal to the argument of the transform.

Performing averaging in Eq.~(\ref{spdf}) over $l$ we find,
\begin{eqnarray}
&& \!\!\!\!\!\!\!\!\!\!\!\! \psi(\tau)=-\frac{d{\tilde p}_v(\tau)}{d\tau},\ \ {\tilde p}_v(\tau)=\int_0^{\infty}\exp\left[-v\tau\right] p_v(v)dv.\label{deriv}
\end{eqnarray}
where ${\tilde p}_v(\tau)$ is the Laplace transform of $p_v(v)$. We have for distribution (\ref{velst}) using integral representation of the incomplete Gamma function,
\begin{eqnarray}
&& \!\!\!\!\!\!\!\!\!\!\!\! {\tilde p}_v(\tau)=\int_0^{\infty}\frac{\exp\left[-v\tau\right] dv}{\Gamma(1-\alpha+1/\alpha)}\int_{v^{\alpha}}^{\infty} \exp[-t]t^{-\alpha}dt.\label{intg}
\end{eqnarray}
In the case of $0<\alpha<1$ we can interchange the order of integrations ($t>v^{\alpha}$ is $v<t^{1/\alpha}$),
\begin{eqnarray}
&& \!\!\!\!\!\!\!\!\!\!\!\! {\tilde p}_v(\tau)=\int_{0}^{\infty} \exp[-t]t^{-\alpha}dt\int_0^{t^{1/\alpha}}\frac{\exp\left[-v\tau\right] dv}{\Gamma(1-\alpha+1/\alpha)}\nonumber\\&&\!\!\!\!\!\!\!\!\!\!\!\!=\frac{\Gamma(1-\alpha)}{\tau\Gamma(1-\alpha+1/\alpha)}-\int_0^{\infty} \frac{\exp[-t-\tau t^{1/\alpha}]t^{-\alpha}dt}{\tau\Gamma(1-\alpha+1/\alpha)}.
\end{eqnarray}
The large $\tau$ asymptotic form of the last integral is found using integration variable $x=\tau t^{1/\alpha}$ so that $t=x^{\alpha}\tau^{-\alpha}$,
\begin{eqnarray}
&& \!\!\!\!\!\!\!\!\!\!\!\! {\tilde p}_v(\tau)=\frac{\Gamma(1-\alpha)}{\tau\Gamma(1-\alpha+1/\alpha)}\nonumber\\&&\!\!\!\!\!\!\!\!\!\!\!\!-\int_0^{\infty} \frac{\alpha\exp[-x^{\alpha}\tau^{-\alpha}-x]x^{-\alpha^2+\alpha-1}\tau^{\alpha^2-\alpha-1}dx}{\Gamma(1-\alpha+1/\alpha)}.
\end{eqnarray}
We can set $x^{\alpha}\tau^{-\alpha}$ in the exponent to zero at large $\tau$ so that we have at $\tau\gg 1$,
\begin{eqnarray}
&& \!\!\!\!\!\!\!\!\!\!\!\! {\tilde p}_v(\tau)\sim \frac{\Gamma(1-\alpha)}{\tau\Gamma(1-\alpha+1/\alpha)}-\frac{\alpha\Gamma(\alpha-\alpha^2)\tau^{\alpha^2-\alpha-1}}{\Gamma(1-\alpha+1/\alpha)}.
\end{eqnarray}
We find that in the leading order the first term dominates and the residence time's PDF has the tail,
\begin{eqnarray}
&& \!\!\!\!\!\!\!\!\!\!\!\! \psi(\tau)\sim \frac{\Gamma(1-\alpha)}{\tau^2\Gamma(1-\alpha+1/\alpha)},\ \  \tau\gg 1,\ \ 0<\alpha<1,
\end{eqnarray}
confirming that the average residence time is infinite, cf. Eq.~(\ref{avrg0}). We stress that in this case there is a universal $\tau^{-2}$ tail where the exponent is independent of $\alpha$ but the prefactor does depend on $\alpha$. In the case of $\alpha=1$ we find using the table of Laplace transforms \cite{Bateman},
\begin{eqnarray}
&& \!\!\!\!\!\!\!\!\!\!\!\! {\tilde p}_v(\tau)=\frac{\ln(\tau+1)}{\tau}, \ \ \psi(\tau)=\frac{(\tau+1)\ln(\tau+1)-\tau}{\tau^2(\tau+1)},\label{lapla}
\end{eqnarray}
so that $\psi(\tau)$ can be written using elementary functions in this case. We find considering large $\tau$ that $\psi(\tau)\sim \tau^{-2}\ln \tau$ with the corresponding divergence of $\langle \tau\rangle$. In the case of $1< \alpha< [\sqrt{5}+1]/2$ we rewrite Eq.~(\ref{intg}) as
\begin{eqnarray}
&& \!\!\!\!\!\!\!\!\!\!\!\! {\tilde p}_v(\tau)=\int_0^{\infty}\frac{\exp\left[-v\tau\right] dv}{(\alpha-1)\Gamma(1-\alpha+1/\alpha)}
\left(\exp[-v^{\alpha}]v^{\alpha(1-\alpha)}\right.\nonumber\\&&\!\!\!\!\!\!\!\!\!\!\!\!\left.-\int_{v^{\alpha}}^{\infty} \exp[-t]t^{-(\alpha-1)}dt\right).\label{inegrals}
\end{eqnarray}
We have,
\begin{eqnarray} && \!\!\!\!\!\!\!\!\!\!\!\! \int_0^{\infty}v^{\alpha(1-\alpha)}\exp\left[-v\tau-v^{\alpha}\right] dv=\tau^{-\alpha(1-\alpha)-1} \nonumber\\&&\!\!\!\!\!\!\!\!\!\!\!\!\times\int_0^{\infty}\!\!x^{\alpha(1\!-\!\alpha)}\exp\left[\!-\!x\!-\!x^{\alpha}\tau^{\!-\!\alpha}\right] dx\!\sim\! \frac{\Gamma(1\!+\!\alpha(1\!-\!\alpha))}{\tau^{1-\alpha(\alpha-1)}},\label{kap}
\end{eqnarray}
where the asymptotic equality holds at large $\tau$. Further we observe that for the last term in Eq.~(\ref{inegrals}) the exponent of $t$ is between $0$ and $1$ so we can use for integration the technique used for $0<\alpha<1$. We find that,
\begin{eqnarray}
&& \!\!\!\!\!\!\!\!\!\!\!\! \int_0^{\infty}\exp\left[-v\tau\right] dv\int_{v^{\alpha}}^{\infty} \exp[-t]t^{1-\alpha}dt\sim \frac{\Gamma(2-\alpha)}{\tau},  \end{eqnarray}
that decays faster than the last term in Eq.~(\ref{kap}). We conclude that the tail of the PDF of the residence time at $1< \alpha< [\sqrt{5}+1]/2$ is described by,
\begin{eqnarray}
&& \!\!\!\!\!\!\!\!\!\!\!\! {\tilde p}_v(\tau)\sim\frac{\Gamma(1-\alpha(\alpha-1))}{(\alpha-1)\Gamma(1-\alpha+1/\alpha)\tau^{1-\alpha(\alpha-1)}},\\&&\!\!\!\!\!\!\!\!\!\!\!\!\psi(\tau)\sim\frac{\Gamma(2-\alpha(\alpha-1))}{(\alpha-1)\Gamma(1-\alpha+1/\alpha)\tau^{2-\alpha(\alpha-1)}}.
\end{eqnarray}
Thus $\psi(\tau)$ has a power law tail. The exponent is continuous at $\alpha=1$ where as $\alpha$ approaches $1$ from above the exponent tends to the value $2$ holding at $\alpha<1$. When $\alpha$ is increased from $1$ to the upper limit $[\sqrt{5}+1]/2$ of the range of physically admissible $\alpha$ the exponent decreases to $1$ so that $\psi(\tau)$ would become non-normalizable at $\alpha=[\sqrt{5}+1]/2$. Thus when $1<\alpha<[\sqrt{5}+1]/2$ we have normalizable power-law tail with divergent average.

We conclude that at $\alpha>1$ the exponent of the power-law is a non-universal, $\alpha-$dependent number. The value of $\alpha=1$ sets the boundary between $\alpha-$independent $\tau^{-2}$ tail and slower decaying tail with $\alpha-$dependent exponent.

We provide the formulas for the tail with restored dimensions that stress dependencies on the experimental parameters. The PDF of the time between consecutive acceleration events obeys,
\begin{eqnarray}
&& \!\!\!\!\!\!\!\!\!\! \psi(\tau)\sim \frac{d}{4(1-\lambda)v_0\tau^2}\frac{\Gamma(1-\alpha)}{\Gamma(1-\alpha+1/\alpha)},\ \ 0<\alpha<1,\\&&\!\!\!\!\!\!\!\!\!\!\psi(\tau)\sim \frac{d}{4(1-\lambda)v_0\tau^2}\ln\left(\frac{4(1-\lambda)v_0\tau}{d}\right),\ \ \alpha=1.
\end{eqnarray}
In the case of $1< \alpha< [\sqrt{5}+1]/2$ we have,
\begin{eqnarray}&&\!\!\!\!\!\!\!\!\!\!
\psi(\tau)\!\sim\!\frac{\Gamma(2-\alpha(\alpha-1))}{(\alpha-1)\Gamma(1-\alpha+1/\alpha)\tau}\left(\frac{d}{4(1-\lambda)v_0\tau}\right)^{1+\alpha-\alpha^2}.\nonumber
\end{eqnarray}
The tail describes $\tau\gg \tau_{typ}$ where $\tau_{typ}= d/[4(1-\lambda)v_0]$ is the typical time between the acceleration events. Below we return to dimensionless variables unless stated otherwise.

The difference between the cases of finite and infinite average residence times can be seen considering the asymptotic growth of the distance $l(t)$ passed by the particle. We have
\begin{eqnarray}&& \!\!\!\!\!\!\!\!\!\!\!\!
\lim_{t\to\infty}\frac{l(t)}{t}=\lim_{t\to\infty}\frac{N(t)}{t}\frac{1}{N(t)}\sum_{k=1}^{N(t)} l_k\label{gr}, \end{eqnarray}
where $N(t)$ is the number of the walk's steps (visited pores) performed by the time $t$. The law of large numbers implies that $\sum_{k=1}^{N} l_k/N$ becomes $\langle l\rangle$ in the limit of large $N$. Further if the average residence time $\langle \tau\rangle$ is finite then $t/N(t)$ becomes $\langle \tau\rangle$ at large times \cite{Feller} so that,
\begin{eqnarray}&& \!\!\!\!\!\!\!\!\!\!\!\!
\lim_{t\to\infty}\frac{l(t)}{t}=\frac{\langle l\rangle}{\langle \tau\rangle}=\left\langle \frac{1}{v}\right\rangle^{-1}. \label{basic}
\end{eqnarray}
where we used Eq.~(\ref{avrg0}). Thus when $\langle \tau\rangle$ is finite (which is the same as $\langle v^{-1}\rangle$ is finite) we have linear asymptotic growth of the passed distance with time. In contrast when $\langle\tau\rangle=\infty$ we find from Eq.~(\ref{gr}) that $\lim_{t\to\infty}l(t)/t=0$ that is the growth is slower than linear (by logarithmic factor see below). This somewhat unusual behavior holds because the particle has small velocity for anomalously long times.

The Laplace transform ${\tilde p}_v(\tau)$ has a simple physical meaning. We consider the probability density function $P_K(t)$ of the number $K$ of new pores visited by the particle in time $t$ (so if the particle stays in the initial pore then $K=0$). We have $K=0$ provided $\tau_0>t$ so that $P_0(t)=\int_t^{\infty}\psi(\tau')d\tau'={\tilde p}_v(t)$ where we used Eq.~(\ref{deriv}). Thus ${\tilde p}_v(t)$ is the probability of not leaving the initial pore during time $t$. This property can be useful in the experimental study of the PDF of velocity (provided the underlying exponential distribution of pore lengths holds).

\section{PDF of the number of channels passed in given time} \label{pd}

In this Section we find the PDF $P_K(t)$ of the number of new channels visited by the particle. Since the length of the channel does not have large fluctuations obeying exponential statistics then $P_K(t)$ is quite similar to the propagator $P(l, t)$ giving the PDF of the distance $l$ passed in time $t$. However, in contrast with the propagator found in next Sections, $P_K(t)$ is simpler for finding which can be useful observation for future studies.

For $K>0$ the particle visits $K$ new pores in time $t$ provided $\sum_{k=0}^{K-1} \tau_k<t$ but $\sum_{k=0}^{K} \tau_k>t$ so that,
\begin{eqnarray}
&& \!\!\!\!\!\!\!\!\!\!\!\!\!\!\! P_K(t)\!=\!\left\langle \int_0^t \!\!dt' \delta\left(\sum_{k=0}^{K\!-\!1} \tau_k\!-\!t'\right)\!\int_t^{\infty}\!\! dt'' \delta\left(\sum_{k=0}^{K} \tau_k\!-\!t''\right)\right\rangle \nonumber\\&&\!\!\!\!\!\!\!\!\!\!\!\!\!\!\!=\!\int_0^t dt'\int_t^{\infty} \!\!dt'' \left\langle  \delta\left(\sum_{k=0}^{K-1} \tau_k-t'\right)\delta\left(\tau_K+t'-t''\right)\right\rangle.
\end{eqnarray}
Since $\tau_k$ are independent then we can average over $\tau_K$ which gives
\begin{eqnarray}
&& \!\!\!\!\!\!\!\!\!\!\!\!\!\!\! P_K(t)\!=\!\int_0^t dt'\int_t^{\infty} dt''\psi(t''-t') \left\langle  \delta\left(\sum_{k=0}^{K-1} \tau_k-t'\right)\right\rangle \nonumber\\&&\!\!\!\!\!\!\!\!\!\!\!\!\!\!\!=\int_0^t dt'{\tilde p}_v(t-t') \left\langle  \delta\left(\sum_{k=0}^{K-1} \tau_k-t'\right)\right\rangle,
\end{eqnarray}
where we used Eq.~(\ref{deriv}). Laplace transform over $t$ gives,
\begin{eqnarray}
&& \!\!\!\!\!\!\!\!\!\!\!\!\!\!\! P_K(s)=\left\langle  \exp[-s\tau]\right\rangle^K\int_0^{\infty}\frac{p_v(v)dv}{s+v},
\end{eqnarray}
where $s$ is the Laplace transform variable, we used independence of $\tau_k$ and
\begin{eqnarray}
&& \!\!\!\!\!\!\!\!\!\!\!\!\!\!\! \int_0^{\infty}\!\!\exp[-s\tau]{\tilde p}_v(\tau)d\tau\!=\!\int_0^{\infty}p_v(v)dv\int_0^{\infty}\!\!\exp[-(s\!+\!v)\tau]d\tau\nonumber\\&&\!\!\!\!\!\!\!\!\!\!\!\!\!\!\!=\int_0^{\infty}\frac{p_v(v)dv}{s+v}=\left\langle \frac{1}{s+v}\right\rangle.
\end{eqnarray}
Finally we observe using ${\tilde p}(\tau=0)=1$,
\begin{eqnarray}
&& \!\!\!\!\!\!\!\!\!\!\!\!\!\!\!  \left\langle  \exp[-s\tau]\right\rangle\!=\!-\int_0^{\infty}\!\!\!\exp[-s\tau]{\tilde p}_v'(\tau)d\tau\!=\!1\!-\!\left\langle \frac{s}{s+v}\right\rangle \nonumber\\&&\!\!\!\!\!\!\!\!\!\!\!\!\!\!\!=\left\langle \frac{v}{s+v}\right\rangle.
\end{eqnarray}
We find thus that,
\begin{eqnarray}
&& \!\!\!\!\!\!\!\!\!\!\!\!\!\!\! P_K(s)=\left\langle \frac{v}{s+v}\right\rangle^K\left\langle \frac{1}{s+v}\right\rangle.\label{fin}
\end{eqnarray}
Though this formula is derived for $K>0$ it is readily checked considering Laplace transform of the previously derived $P_0(t)$ that it holds for $K=0$ as well. The normalization condition $\sum_{K=0}^{\infty}P_K(s)=s^{-1}$ is readily checked. We conclude that the distribution of the number of new pores visited by the particle in time $t$ is,
\begin{eqnarray}
&& \!\!\!\!\!\!\!\!\!\!\!\!\!\!\! P_K(t)\!=\!\!\int_{\epsilon-i\infty}^{\epsilon+i\infty} \!\!\frac{\exp[st]ds}{2\pi i}\left\langle \frac{v}{s+v}\right\rangle^K\!\left\langle \frac{1}{s+v}\right\rangle.\label{fint}
\end{eqnarray}
This formula relies on the exponential distribution of pore lengths and can be used for arbitrary distribution of velocity. For distribution (\ref{velst}) we find
\begin{eqnarray}&&\!\!\!\!\!\!\!\!\!\!\!\!\left\langle \frac{1}{s+v}\right\rangle =\int_0^{\infty}\frac{(s+v)^{-1} dv}{\Gamma(1-\alpha+1/\alpha)}\int_{v^{\alpha}}^{\infty} \exp[-t]t^{-\alpha}dt \nonumber\\&&\!\!\!\!\!\!\!\!\!\!\!\!\!=\!\int_0^{\infty}dt\exp[-t]t^{-\alpha}\int_0^{t^{1/\alpha}}\frac{(s+v)^{-1} dv}{\Gamma(1-\alpha+1/\alpha)}\nonumber\\&&\!\!\!\!\!\!\!\!\!\!\!\!
\!=\!\int_0^{\infty}\!\!\frac{\exp[-t]t^{-\alpha}\left[\alpha^{-1}\ln t\!-\!\ln s\!+\!\ln(1\!+\!st^{-1/\alpha})\right]dt}{\Gamma(1\!-\!\alpha\!+\!1/\alpha)},\label{integro}\end{eqnarray}
that for $\alpha=1$ reduces to the definition of $\langle (s+v)^{-1}\rangle$ as Laplace transform of ${\tilde p}_v(\tau)$ given by Eq.~(\ref{lapla}),
\begin{eqnarray}&&\!\!\!\!\!\!\!\!\!\!\!\!\left\langle \frac{1}{s+v}\right\rangle 
=\int_0^{\infty}\frac{\exp[-st]\ln(1+t)dt}{t}.\label{intgral}
\end{eqnarray}
It seems impossible writing $P_K(t)$ via elementary functions but we can use it for studying moments. We find for Laplace transform of the average number of new pores visited in time $t$,
\begin{eqnarray}
&& \!\!\!\!\!\!\!\!\!\!\!\!\!\!\!\langle K\rangle(s)=\sum_{K=0}^{\infty} K P_K(s)=s^{-2}\left\langle \frac{1}{s+v}\right\rangle^{-1}\!-s^{-1}.\label{rf}
\end{eqnarray}
where we used Eq.(\ref{fin}) and $\sum_k kx^k=x(1-x)^{-2}$. Similar formulas can be written for higher-order moments.

We consider the long-time behavior of $\langle K(t)\rangle$ that is determined by the small $s$ behavior of $\langle K\rangle(s)$. In models with finite $\langle v^{-1}\rangle$ we have at small $s$,
\begin{eqnarray}
&& \!\!\!\!\!\!\!\!\!\!\!\!\!\!\!\langle K\rangle(s)=s^{-2}\left\langle \frac{1}{v}\right\rangle^{-1}=\frac{1}{s^2\langle \tau\rangle}.\end{eqnarray}

We find the long-time behavior,
\begin{eqnarray}
&& \!\!\!\!\!\!\!\!\!\!\!\!\!\!\!\langle K(t)\rangle=\frac{t}{\langle \tau\rangle},\label{averagen}\end{eqnarray}
that reproduces the law of large numbers for $N(t)/t$ described previously. In the model (\ref{velst}) with infinite $\langle \tau\rangle$ the behavior is less intuitive. We start from the case of $\alpha<1$ where we find from Eq.~(\ref{integro}) that for $\alpha<1$,
\begin{eqnarray}&&\!\!\!\!\!\!\!\!\!\!\!\!\!\!\!\!\!\left\langle\! \frac{1}{s\!+\!v}\!\right\rangle
\!=\!\frac{\Gamma(1\!-\!\alpha)\ln(1/s)}{\Gamma(1\!-\!\alpha\!+\!1/\alpha)} \!+\!\frac{\psi_{\Gamma}(1\!-\!\!\alpha)\Gamma(1\!-\!\alpha)}{\alpha\Gamma(1\!-\!\alpha\!+\!1/\alpha)}\!+\!O(s),\label{inv1}
\end{eqnarray}
where $\psi_{\Gamma}(x)=\Gamma'(x)/\Gamma(x)$ is the digamma function. This can be confirmed directly from Eq.~(\ref{as}) using logarithmic divergence of $\langle v^{-1}\rangle$ at small velocities. Thus
\begin{eqnarray}&&\!\!\!\!\!\!\!\!\!\!\!\!\!\!\!\left\langle \frac{1}{s\!+\!v}\right\rangle^{-1}\!\!\!\!\sim\!\frac{\Gamma(1\!-\!\alpha\!+\!1/\alpha)}{\Gamma(1\!-\!\alpha)\ln (1/s)}\left[1\!-\!\frac{\psi(1\!-\!\!\alpha)}{\alpha \ln (1/s)}\right],\ \alpha<1, \label{sm} \end{eqnarray}
which can be confirmed directly from Eq.~(\ref{as}) using logarithmic divergence of $\langle v^{-1}\rangle$ at small velocities. We conclude from Eq.~(\ref{rf}) that $\langle K\rangle(s)$ at small $s$ is,
\begin{eqnarray}
&& \!\!\!\!\!\!\!\!\!\!\!\!\!\!\!\langle K\rangle(s)=-\frac{\Gamma(1-\alpha+1/\alpha)}{\Gamma(1-\alpha)s^{2}\ln s}-\!\frac{\Gamma(1-\alpha+1/\alpha)\psi_{\Gamma}(1\!-\!\!\alpha)}{\alpha\Gamma(1-\alpha) s^{2}\ln^2 s}
\nonumber\\&&\!\!\!\!\!\!\!\!\!\!\!\!\!\!\!
+O\left(\frac{1}{s^2\ln^3 s}\right).\label{asympto}
\end{eqnarray}
Thus the long-time behavior of $\langle K(t)\rangle$ obeys,
\begin{eqnarray}
&& \!\!\!\!\!\!\!\!\!\!\!\!\!\!\!\langle K(t)\rangle\sim-\frac{\Gamma(1-\alpha+1/\alpha)}{\Gamma(1-\alpha)}\int_{\epsilon-i\infty}^{\epsilon+i\infty} \frac{ds}{2\pi i}\frac{\exp[st]}{s^{2}\ln s}\nonumber\\&&\!\!\!\!\!\!\!\!\!\!\!\!\!\!\!\times\left[1+\frac{\psi_{\Gamma}(1\!-\!\!\alpha)}{\alpha\ln s}\right]+O\left(\frac{t}{\ln^3 t}\right),\label{lpl}
\end{eqnarray}
where the inverse Laplace transform of $1/[s^2\ln^3 s]$ is proportional to $t/\ln^3 t$, see below.
We stress that the integral in the RHS is not inverse Laplace transform of $1/s^{2}\ln s$ because $1/s^{2}\ln s$ has simple pole at $s=1$ (the residue there is not counted see \cite{Hardy}). We have rescaling integration variable by $t$ (we use the same letter for integration variable with no ambiguity and rescale the infinitesimal $\epsilon$ correspondingly) for the first integral,
\begin{eqnarray}
&& \!\!\!\!\!\!\!\!\!\!\!-\int_{\epsilon-i\infty}^{\epsilon+i\infty} \frac{ds}{2\pi i}\frac{\exp[st]}{s^{2}\ln s}=\frac{t}{\ln t}\int_{\epsilon-i\infty}^{\epsilon+i\infty} \frac{ds}{2\pi i}\frac{\exp[s]}{s^{2}[1-\ln s/\ln t]}\nonumber\\&&\!\!\!\!\!\!\!\!\!\!\!=
\frac{t}{\ln t}+\frac{t(1-C)}{\ln^2 t}+O\left(\frac{t}{\ln^3 t}\right),
\end{eqnarray}
where we performed expansion of the integrand in $\ln s/\ln t$ and used inverse Laplace transform of $\ln s/s^2$ from \cite{Bateman} with $C$ Euler's constant. We find from Eq.~(\ref{lpl})
\begin{eqnarray}
&& \!\!\!\!\!\!\!\!\langle K(t)\rangle\!\sim\!\frac{\Gamma(1\!-\!\alpha\!+\!1/\alpha)}{\Gamma(1\!-\!\alpha)}\left[\frac{t}{\ln t}\!+\!\frac{t}{\ln^2 t}\left(1\!-\!C\!-\!\frac{\psi_{\Gamma}(1\!-\!\!\alpha)}{\alpha}\right)\right]\nonumber\\&&\!\!\!\!\!\!\!\!+O\left(\frac{t}{\ln^3 t}\right),\ \ \alpha<1. \label{sml}
\end{eqnarray}
In the leading order we have $\langle K(t)\rangle\propto t/\ln t$. Thus the average number of pores visited by the particle in time $t$ grows slower than linearly with time. This is the manifestation of the logarithmic divergence of $\langle \tau\rangle$ at large $\tau$, cf. Eq.~(\ref{averagen}).

In $\alpha=1$ case the behavior of $\langle K(t)\rangle$ resembles the $\alpha<1$ case. The asymptotic form of $\left\langle [s+v]^{-1}\right\rangle$ at small $s$ is found observing that Eq.~(\ref{intgral}) implies,
\begin{eqnarray}&&\!\!\!\!\!\!\!\!\!\!\!\!\frac{d}{ds}\left\langle \frac{1}{s+v}\right\rangle=-\int_0^{\infty}\exp[-st]\ln(1+t)dt\nonumber\\&&\!\!\!\!\!\!\!\!\!\!\!\!\!\!\!=\frac{\exp[s]}{s}Ei(-s)\sim \frac{\ln s}{s}+\frac{C}{s}+\ldots,
\end{eqnarray}
where we used the integral from \cite{Gradshteyn} and the asymptotic equality is written at small $s$. We conclude that (integration constant is negligible),
\begin{eqnarray}&&\!\!\!\!\!\!\!\!\!\!\!\!
\left\langle \frac{1}{s+v}\right\rangle\sim \frac{\ln^2 s}{2}+C\ln s.
\end{eqnarray}
Comparing this with Eq.~(\ref{inv1}) we see that divergence of $\Gamma(1-\alpha)$ at $\alpha=1$ brings more singular behavior of the average at small $s$.
We have,
\begin{eqnarray}&&\!\!\!\!\!\!\!\!\!\!\!\!
\left\langle \frac{1}{s+v}\right\rangle^{-1}\sim \frac{2}{\ln^2 s}-\frac{4C}{\ln^3 s},\ \ \alpha=1. \label{inv2}
\end{eqnarray}
We find for the long-time behavior of $\langle K(t)\rangle$,
\begin{eqnarray}
&& \!\!\!\!\!\!\!\!\!\!\!\!\!\!\!\langle K(t)\rangle\sim 2\int_{\epsilon-i\infty}^{\epsilon+i\infty} \frac{ds}{2\pi i}\frac{\exp[st]}{s^{2}\ln^2 s}\left[1 -\frac{2C}{\ln s}\right],\ \ \alpha=1,\label{avrgrowtho}
\end{eqnarray}
with correction of order $t/\ln^4 t$. We use that,
\begin{eqnarray}
&& \!\!\!\!\!\!\!\!\!\!\!\!\!\!\!\int_{\epsilon-i\infty}^{\epsilon+i\infty} \frac{ds}{2\pi i}\frac{\exp[st]}{s^{2}\ln^2 s}=\frac{t}{\ln^2 t}\int_{\epsilon-i\infty}^{\epsilon+i\infty} \frac{ds}{2\pi i}\frac{\exp[s]}{s^{2}[1-\ln s/\ln t]^2}\nonumber\\&&\!\!\!\!\!\!\!\!\!\!\!\!\!\!\!=\frac{t}{\ln^2 t}+\frac{2(1-C)t}{\ln^3 t}+O\left(\frac{t}{\ln^4 t}\right).
\end{eqnarray}
We find from Eq.~(\ref{avrgrowtho}) that,
\begin{eqnarray}
&& \!\!\!\!\!\!\!\!\!\!\!\!\!\!\!\langle K(t)\rangle\sim \frac{2t}{\ln^2 t}+\frac{4t}{\ln^3 t}+O\left(\frac{t}{\ln^4 t}\right),\ \ \alpha=1.\label{avrgrowth}
\end{eqnarray}
Thus the leading order growth $t/\ln^2 t$ is logarithmically slower than in $\alpha<1$ case, cf. Eq.~(\ref{sml}).

In the performed calculations we kept the next order term beyond the leading order one. This is because this is needed for the calculation of dispersion below and because the series is in $1/\ln t$ where $\ln t$ is never too large so higher order corrections can be necessary for comparison with observations. In $\alpha>1$ case the situation is different and there is no need for going beyond the leading order calculation. Using $t^{-\alpha}=[t^{1-\alpha}]'/(1-\alpha)$ we write Eq.~(\ref{integro}) in the form,
\begin{eqnarray}&&\!\!\!\!\!\!\!\!\!\!\!\!\left\langle \frac{1}{s+v}\right\rangle=\int_0^{\infty}\frac{\exp[-t]t^{1-\alpha}\ln(1+t^{1/\alpha}/s)dt}{(1-\alpha)\Gamma(1-\alpha+1/\alpha)}\nonumber\\&&\!\!\!\!\!\!\!\!\!\!\!\!-\int_0^{\infty}\frac{\exp[-t]t^{1/\alpha-\alpha} dt}{\alpha(s+t^{1/\alpha})(1-\alpha)\Gamma(1-\alpha+1/\alpha)}\label{rf0}.\end{eqnarray} The first integral has logarithmic divergence at small $s$ but the second diverges as a power-law. When $s$ is small that integral is determined by small $t$. Introducing infinitesimal fixed parameter $\epsilon$ (that disappears from the final answer) we can write this integral asymptotically as,
\begin{eqnarray}&&\!\!\!\!\!\!\!\!\!\!\!\!\int_0^{\epsilon}\frac{t^{1/\alpha-\alpha} dt}{\alpha(s+t^{1/\alpha})}=s^{\alpha-\alpha^2}\int_0^{\epsilon^{1/\alpha}/s}\frac{x^{\alpha-\alpha^2} dx}{1+x}\sim s^{\alpha-\alpha^2}\nonumber\\&&\!\!\!\!\!\!\!\!\!\!\!\! \times\int_0^{\infty}\frac{x^{\alpha-\alpha^2} dx}{1+x}=s^{\alpha-\alpha^2} \Gamma(1+\alpha-\alpha^2)\Gamma(\alpha^2-\alpha).\end{eqnarray}
We conclude that,
\begin{eqnarray}&&\!\!\!\!\!\!\!\!\!\!\!\!\left\langle \frac{1}{s+v}\right\rangle\sim \frac{s^{\alpha-\alpha^2}\alpha|\Gamma(\alpha-\alpha^2)|\Gamma(\alpha^2-\alpha)}{\Gamma(1-\alpha+1/\alpha)},\ \ \alpha>1. \label{inv3}\end{eqnarray}
Combining this with Eqs.~(\ref{rf}),(\ref{rf0}) we find,
\begin{eqnarray}&&\!\!\!\!\!\!\!\!\!\!\!\!\left\langle K(s)\right\rangle\sim
\frac{\Gamma(1-\alpha+1/\alpha)}{s^{\alpha-\alpha^2+2}\alpha|\Gamma(\alpha-\alpha^2)|\Gamma(\alpha^2-\alpha)},\\&&\!\!\!\!\!\!\!\!\!\!\!\!\left\langle K(t)\right\rangle\!\sim\!\frac{t^{1+\alpha-\alpha^2}\Gamma(1\!-\!\alpha\!+\!1/\alpha)}{\Gamma(\alpha\!-\!\alpha^2\!+\!2)\alpha|\Gamma(\alpha\!-\!\alpha^2)|\Gamma(\alpha^2\!-\!\alpha)},\ \ \!\alpha\!>\!1.\label{avrg}
\end{eqnarray}
We find the power law growth with $\alpha-$dependent exponent. This is in contrast with $\alpha<1$ where the law of growth is $\alpha-$independent. The growth law's exponent decreases monotonously when $\alpha$ increases from $1$ to $[\sqrt{5}+1]/2$. We have linear growth in time of the average number of visited pores when $\alpha\to 1$. In contrast the exponent tends to zero when $\alpha\to [\sqrt{5}+1]/2$ so that in this limit the average number of visited pores almost does not grow with time. This is because of high probability of very low velocities - the PDF has integrable singularity at $v=0$ becoming non-integrable in $\alpha\to [\sqrt{5}+1]/2$ limit.

We consider the dispersion of the number of visited pores determined by the second moment,
\begin{eqnarray}
&& \!\!\!\!\!\!\!\!\!\!\!\!\!\!\!\langle K^2\rangle(s)=\sum_{K=0}^{\infty} K^2 P_K(s)=\frac{2}{s^3}\left\langle \frac{1}{s+v}\right\rangle^{-2}\nonumber\\&&\!\!\!\!\!\!\!\!\!\!\!\!\!\!\!-\frac{3}{s^2}\left\langle \frac{1}{s+v}\right\rangle^{-1}+\frac{1}{s},\label{quadr}
\end{eqnarray}
where we used
\begin{eqnarray}
&& \!\!\!\!\!\!\!\!\!\!\!\!\!\!\!\sum_{k=0}^{\infty} k^2 x^k=\sum_{k=0}^{\infty} k(k-1) x^k+\frac{x}{(1-x)^2}=\frac{x(x+1)}{(1-x)^3}.
\end{eqnarray}
It is seen readily from Eqs.~(\ref{sm}), (\ref{inv2}) and (\ref{inv3}) that for all $\alpha$ the small-$s$ divergence of the first term in Eq.~(\ref{quadr}) is stronger than that of the rest of terms so that,
\begin{eqnarray} && \!\!\!\!\!\!\!\!\!\!\!\!\!\!\!\langle K^2\rangle(s)\sim \frac{2}{s^3}\left\langle \frac{1}{s+v}\right\rangle^{-2}.\label{disp}\end{eqnarray}
We find from Eqs.~(\ref{sm}), (\ref{inv2}) and (\ref{inv3}) that the leading order behavior at small $s$ is,
\begin{eqnarray} && \!\!\!\!\!\!\!\!\!\!\!\!\!\!\!\langle K^2\rangle(s)\sim \frac{2\Gamma^2(1\!-\!\alpha\!+\!1/\alpha)}{\Gamma^2(1\!-\!\alpha)s^3\ln^2 s},\ \ \alpha<1\\&&\!\!\!\!\!\!\!\!\!\!\!\!\!\!\!\langle K^2\rangle(s)\sim \frac{8}{s^3\ln^4 s},\ \ \alpha=1.
\end{eqnarray}
In the case of $\alpha>1$ we find,
\begin{eqnarray} && \!\!\!\!\!\!\!\!\!\!\!\!\!\!\!\langle K^2\rangle(s)\sim \frac{2\Gamma^2(1-\alpha+1/\alpha)}{s^{2\alpha-2\alpha^2+3}\alpha^2\Gamma^2(\alpha-\alpha^2)\Gamma^2(\alpha^2-\alpha)}.\end{eqnarray}
For finding the corresponding time dependencies we observe that rescaling the integration variable in the inverse Laplace transforms puts in the integrands $\ln t$ instead of $\ln s$ in the leading order at large $t$. We find,
\begin{eqnarray} && \!\!\!\!\!\!\!\!\!\!\!\!\!\!\!\langle K^2(t)\rangle\sim \frac{t^2\Gamma^2(1\!-\!\alpha\!+\!1/\alpha)}{\Gamma^2(1\!-\!\alpha)\ln^2 t}\sim \langle K(t)\rangle^2,\ \ \alpha<1, \label{smaller0}\\&&\!\!\!\!\!\!\!\!\!\!\!\!\!\!\!\langle K^2(t)\rangle\sim \frac{4t^2}{\ln^4 t}\sim \langle K(t)\rangle^2,\ \ \alpha=1, \label{smaller}\\&&\!\!\!\!\!\!\!\!\!\!\!\!\!\!\!\langle K^2(t)\rangle\sim \frac{2t^{2\alpha-2\alpha^2+2}\Gamma^2(1-\alpha+1/\alpha)}{\alpha^2\Gamma^2(\alpha-\alpha^2)\Gamma^2(\alpha^2-\alpha)\Gamma(2\alpha-2\alpha^2+3)}\nonumber\\&&\!\!\!\!\!\!\!\!\!\!\!\!\!\!\!\sim \frac{2\Gamma^2(\alpha\!-\!\alpha^2\!+\!2)\langle K(t)\rangle^2}{\Gamma(2\alpha-2\alpha^2+3)},\ \ \alpha>1.\end{eqnarray}
We observe that in the case of $\alpha>1$ both the second moment and dispersion of $K(t)$ grow as power-law in time. That law changes from $t^2$ growth at $\alpha$ close to one to linear growth in time at $\alpha=[\sqrt{3}+1]/2$ (superdiffusion). When $\alpha$ increases from $[\sqrt{3}+1]/2$ to $[\sqrt{5}+1]/2$ the power law exponent decreases from one to zero (subdiffusion).

We find from Eqs.~(\ref{smaller0})-(\ref{smaller}) that finding the dispersion $\sigma^2(t)=\langle K^2(t)\rangle-\langle K(t)\rangle^2$ at $\alpha\leq 1$ demands next order corrections. Thus
\begin{eqnarray} && \!\!\!\!\!\!\!\!\!\!\!\!\!\!\! \lim_{t\to\infty}\frac{\langle K^2(t)\rangle-\langle K(t)\rangle^2}{\langle K(t)\rangle^2}=0,\ \ \alpha\leq 1.\label{limit}\end{eqnarray}
In contrast, for $\alpha>1$, we find
\begin{eqnarray} && \!\!\!\!\!\!\!\!\!\!\!\!\!\!\! \lim_{t\to\infty}\frac{\langle K^2(t)\rangle-\langle K(t)\rangle^2}{\langle K(t)\rangle^2}=\frac{2\Gamma^2(\alpha\!-\!\alpha^2\!+\!2)}{\Gamma(2\alpha-2\alpha^2+3)},\ \ \alpha>1.\end{eqnarray}
We consider finding the leading order behavior of $\sigma^2(t)$ at large $t$. For $\alpha=1$ using Eq.~(\ref{inv2}),
\begin{eqnarray} && \!\!\!\!\!\!\!\!\!\!\!\!\!\!\!\langle K^2\rangle(s)\sim \frac{2}{s^3}\left[\frac{2}{\ln^2 s}-\frac{4C}{\ln^3 s}\right]^2\sim\frac{8}{s^3\ln^4 s}-\frac{32C}{s^3\ln^5s}. \end{eqnarray}
We use that,
\begin{eqnarray}
&& \!\!\!\!\!\!\!\!\!\!\!\!\!\!\!\int_{\epsilon-i\infty}^{\epsilon+i\infty} \frac{ds}{\pi i}\frac{\exp[st]}{s^{3}\ln^4 s}=\frac{2t^2}{\ln^4 t}\int_{\epsilon-i\infty}^{\epsilon+i\infty} \frac{ds}{2\pi i}\frac{\exp[s]}{s^{3}[1-\ln s/\ln t]^4}\\&&\!\!\!\!\!\!\!\!\!\!\!\!\!\!\!=\frac{t^2}{\ln^4 t}+\frac{2(3-2C)t^2}{\ln^5 t}+O\left(\frac{t^2}{\ln^6 t}\right),
\end{eqnarray}
where we used the inverse Laplace transform of $\ln s/s^3$ from \cite{Bateman}. Thus
\begin{eqnarray} && \!\!\!\!\!\!\!\!\!\!\!\!\!\!\!\langle K^2(t)\rangle= \frac{4t^2}{\ln^4 t}+\frac{24t^2}{\ln^5 t}+O\left(\frac{t^2}{\ln^6 t}\right).\end{eqnarray}
We conclude using Eq.~(\ref{avrgrowth}) that,
\begin{eqnarray} && \!\!\!\!\!\!\!\!\!\!\!\!\!\!\!\langle K^2(t)\rangle-\langle K(t)\rangle^2=\frac{8t^2}{\ln^5 t}+O\left(\frac{t^2}{\ln^6 t}\right),\ \  \alpha=1.\end{eqnarray}
Thus the normalized dispersion obeys,
\begin{eqnarray} && \!\!\!\!\!\!\!\!\!\!\!\!\!\!\!\frac{\langle K^2(t)\rangle-\langle K(t)\rangle^2}{\langle K(t)\rangle^2}\sim \frac{2}{\ln t},\ \  \alpha=1,\end{eqnarray}
providing details on the limit (\ref{limit}) at $\alpha=1$. In the case $\alpha<1$ using Eq.~(\ref{sm})
\begin{eqnarray}&&\!\!\!\!\!\!\!\!\!\!\!\!\langle K^2\rangle(s)\!\sim\!\frac{2\Gamma^2(1\!-\!\alpha\!+\!1/\alpha)}{\Gamma^2(1\!-\!\alpha)s^3\ln^2 s}+\!\frac{4\psi_{\Gamma}(1\!-\!\!\alpha)\Gamma^2(1\!-\!\alpha\!+\!1/\alpha)}{\alpha \Gamma^2(1\!-\!\alpha)s^3\ln^3 s}.\nonumber
\end{eqnarray}
We use,
\begin{eqnarray}
&& \!\!\!\!\!\!\!\!\!\!\!\!\!\!\!\int_{\epsilon-i\infty}^{\epsilon+i\infty} \frac{ds}{\pi i}\frac{\exp[st]}{s^{3}\ln^2 s}=\frac{t^2}{\ln^2 t}\int_{\epsilon-i\infty}^{\epsilon+i\infty} \frac{ds}{\pi i}\frac{\exp[s]}{s^{3}[1-\ln s/\ln t]^2}\nonumber\\&&\!\!\!\!\!\!\!\!\!\!\!\!\!\!\!=\frac{t^2}{\ln^2 t}+\frac{(3-2C)t^2}{\ln^3 t}+O\left(\frac{t^2}{\ln^4 t}\right).
\end{eqnarray}
We find that,
\begin{eqnarray}&&\!\!\!\!\!\!\!\!\!\!\!\!\langle K^2(t)\rangle\!=\!\frac{\Gamma^2(1\!-\!\alpha\!+\!1/\alpha)}{\Gamma^2(1\!-\!\alpha)}\left[\frac{t^2}{\ln^2 t}+\frac{t^2}{\ln^3 t}\left(3-2C\right.\right.\nonumber\\&&\!\!\!\!\!\!\!\!\!\!\!\!\!\!\!\left.\left.+\frac{2\psi_{\Gamma}(1\!-\!\!\alpha)}{\alpha}\right)
\right]+O\left(\frac{t^2}{\ln^4 t}\right).
\end{eqnarray}
We find using Eq.~(\ref{sml}) that,
\begin{eqnarray}&&\!\!\!\!\!\!\!\!\!\!\!\!\!\sigma^2=\!\frac{\Gamma^2(1\!-\!\alpha\!+\!1/\alpha)t^2}{\Gamma^2(1\!-\!\alpha)\ln^3 t}\left(1\!+\!\frac{4\psi_{\Gamma}(1\!-\!\!\alpha)}{\alpha}\right).
\end{eqnarray}
Thus the normalized dispersion obeys,  \begin{eqnarray} && \!\!\!\!\!\!\!\!\!\!\!\!\!\!\!\frac{\langle K^2(t)\rangle\!-\!\langle K(t)\rangle^2}{\langle K(t)\rangle^2}\sim \frac{1}{\ln t}\left(1+\!\frac{4\psi_{\Gamma}(1\!-\!\!\alpha)}{\alpha}\right),\ \  \alpha<1,\end{eqnarray} providing the description of the limit (\ref{limit}) in $\alpha<1$ case.

The law (\ref{limit}) tells that the variance of $K(t)/\langle K(t)\rangle$ is zero in $t\to\infty$ limit. Thus using Chebichev's inequality we find that the limit in probability holds,
\begin{eqnarray} && \!\!\!\!\!\!\!\!\!\!\!\!\!\!\! \lim_{t\to\infty}\frac{K(t)}{\langle K(t)\rangle}=1,\ \ \alpha\leq 1,\label{limitstrong}\end{eqnarray}
In other words the probability of finite deviations of $K(t)/\langle K(t)\rangle$ from $1$ decays to zero at large times.

We conclude that there is a qualitative difference between $\alpha\leq 1$ and $\alpha>1$ cases. In $\alpha>1$ case both the first and the second moment of the number of visited pores grow as a power law without logarithmic corrections. Dispersion of the passed distance grows proportionally to the square of the mean passed distance. The growth law describes superdiffusion at $1<\alpha<[\sqrt{3}+1]/2$ and subdiffusion at $[\sqrt{3}+1]/2<\alpha<[\sqrt{5}+1]/2$. The variable $K(t)/\langle K(t)\rangle$ has finite fluctuations of order one (It is plausible that the PDF of $K(t)/\langle K(t)\rangle$ has finite $t\to\infty$ limit in this case. The study of this question can be done on the basis of Eq.~(\ref{fint}) but is beyond the scope of this work). In sharp contrast at $\alpha<1$ dispersion grows slower that the square of the mean passed distance implying that the probability of fluctuations of $K(t)/\langle K(t)\rangle$ decays in $t\to\infty$ limit so that Eq.~(\ref{limitstrong}) holds. The dispersion grows ballistically (quadratically in time) with logarithmic correction. Up to that correction this growth is quite similar to L\'{e}vy walk with infinite $\langle \tau\rangle$. This similarity does not hold for $\alpha>1$.

\section{Anomalous diffusion} \label{dif}

In this Section we find the first and second moments of $l(t)$ in the limit of large times. Since in our CTRW the PDF of $l(t)$ vanishes for $l<0$ then it is useful to use the double Laplace transform in $t$ and $l$ which is real function of its arguments,
\begin{eqnarray}&& \!\!\!\!\!\!\!\!\!\!\!\!
P(s, p)=\int_0^{\infty} dt\int_0^{\infty} dl \exp[-st-pl]P(t, l),
\end{eqnarray}
where $P(t, l)$ is the PDF of $l(t)$. We use that $P(s, p)=p(s, -ip)$ where $p(s, k)$ is provided by the Montroll-Weiss equation (\ref{os}). We observe that in our model the joint PDF of $\tau$ and $l$ is given by $\psi'(\tau, v)=\exp[-v\tau]vp_v(v)$. This implies $\psi'(s, v)=vp_v(v)/[s+v]$ which use in Eq.~(\ref{os}) gives,
\begin{eqnarray}
&& \!\!\!\!\!\!\!\!\!\!\!\! P(s, p)=\left\langle\frac{1}{s+(1+p)v}\right\rangle\left\langle\frac{s+pv}{s+(1+p)v}\right\rangle^{-1}.\label{1dlv}
\end{eqnarray}
Here and below the averages are over statistics of velocity $v$. The moments are obtained by differentiating $P(s, p)$.
We find for the Laplace transform of the average distance,
\begin{eqnarray}&& \!\!\!\!\!\!\!\!\!\!\!\!
\langle l\rangle(s)=-\frac{\partial P(s, p)}{\partial p}(p=0)=\frac{\omega(s)}{s^2},\label{avrd}
\end{eqnarray}
where we introduced frequency dependent velocity,
\begin{eqnarray}&& \!\!\!\!\!\!\!\!\!\!\!\!
\omega(s)= 
\left\langle \frac{1}{s\!+\!v}\right \rangle^{-1}\!\left[1\!-\!\left\langle\frac{s}{s\!+\!v}\right \rangle\right].\label{rev}
\end{eqnarray}
If the particle velocity would not fluctuate, $v=v_0$ then we would find $\langle l\rangle(s)=v_0s^{-2}$ recovering Laplace transform of $\langle l\rangle(t)=v_0t$ (here we restore dimensions for clarity). However if there are finite fluctuations of velocity then the law $\langle l\rangle(t)=\langle v\rangle t$ that could be thought valid does not hold. If the velocity has finite moment of order $-1$ then the small $s$ behavior of $\omega(s)$ and the corresponding long-time behavior of $\langle l(t)\rangle$ are described respectively with,
\begin{eqnarray}&& \!\!\!\!\!\!\!\!\!\!\!\!
\omega(s)\sim\left\langle \frac{1}{v}\right \rangle^{-1},\ \ \langle l(t)\rangle\sim t\left\langle \frac{1}{v}\right \rangle^{-1},
\end{eqnarray}
in agreement with Eq.~(\ref{basic}). In the case of divergent $\langle v^{-1}\rangle$ the growth of $\langle l(t)\rangle$ is not linear in time. We consider the model in the long-time limit. It is seen readily from the asymptotic forms of $\langle [s+v]^{-1}\rangle$ given by Eqs.~(\ref{sm}), (\ref{inv2}) and (\ref{inv3}) that independently of $\alpha$ we have
\begin{eqnarray}&& \!\!\!\!\!\!\!\!\!\!\!\!
1\!-\!\left\langle\frac{s}{s\!+\!v}\right \rangle=1+o(s),
\end{eqnarray}
implying by Eq.~(\ref{rev}) that $\omega(s)\sim \left\langle [s\!+\!v]^{-1}\right \rangle^{-1}$ at small $s$. We find using this in Eq.~(\ref{avrd}) and comparing with Eq.~(\ref{rf}) that in the long-time limit,
\begin{eqnarray}&& \!\!\!\!\!\!\!\!\!\!\!\!
\langle l(t)\rangle\sim \langle K(t)\rangle.\label{symp}
\end{eqnarray}
Thus the average distance passed becomes asymptotically equal to the average number of visited pores. This agrees with intuitive formula $\langle l(t)\rangle=\langle K(t)\rangle\langle l \rangle$ (the average passed distance is the average number of the passed pores times the average length of the pore) using that the average length of the pore $\langle l\rangle$ is $1$. This relation holds because strong fluctuations of $l$ have exponentially small probability so setting $l$ equal to its average is valid.

The second moment of the distance passed by the particle is found,
\begin{eqnarray}&& \!\!\!\!\!\!\!\!\!\!\!\!
\langle l^2\rangle(s)=\frac{\partial^2 P(s, p)}{\partial p^2}(p=0)\nonumber\\&&\!\!\!\!\!\!\!\!\!\!\!\!\!\!\!=2s^{-2}\left\langle \frac{1}{s+v}\right \rangle^{-1}\!\left\langle\frac{v^2}{[s\!+ v]^2}\right \rangle\left[1+\frac{\omega(s)}{s}\right].
\end{eqnarray}
We observe that,
\begin{eqnarray}&& \!\!\!\!\!\!\!\!\!\!\!\!
\left\langle\frac{v^2}{[s\!+ v]^2}\right \rangle=1-s^2\left\langle \frac{1}{(s+v)^2}\right\rangle-2s\left\langle \frac{v}{(s+v)^2}\right\rangle\nonumber\\&&\!\!\!\!\!\!\!\!\!\!\!\!=1-\frac{d}{ds}\left\langle \frac{s^2}{s+v}\right\rangle.
\end{eqnarray}
Using the small $s$ behavior of $\langle [s+v]^{-1}\rangle$ given by Eqs.~(\ref{inv1}), (\ref{inv2}) and (\ref{inv3}) we find that the last term vanishes at $s=0$ so that independently of $\alpha$,
\begin{eqnarray}&& \!\!\!\!\!\!\!\!\!\!\!\!
\left\langle\frac{v^2}{[s\!+ v]^2}\right \rangle\sim 1.
\end{eqnarray}
Thus using the small $s$ behavior of $\omega(s)$ we can write,
\begin{eqnarray}&& \!\!\!\!\!\!\!\!\!\!\!\!
\langle l^2\rangle(s)\sim \frac{2}{s^3}\left\langle \frac{1}{s+v}\right \rangle^{-2},
\end{eqnarray}
that holds independently of $\alpha$. Comparing this with Eq.~(\ref{disp}) we find,
\begin{eqnarray}&& \!\!\!\!\!\!\!\!\!\!\!\!
\langle l^2(t)\rangle\sim \langle K^2(t)\rangle.
\end{eqnarray}
This agrees with the $\langle l^2\rangle=\langle K^2\rangle\langle l^2\rangle$ where $\langle l^2\rangle=1$, cf. the discussion of Eq.~(\ref{symp}). Thus at least as far as the large-time growth of the first two moments of $l(t)$ is concerned we can use $K(t)$ instead of $l(t)$ and transfer the results from Section \ref{pd}. This is done in the last Section.

\section{Discussion}

In this Section we summarize the main results and discuss their implications. We also provide the average and dispersion of $l(t)$ which are direct consequences of the considerations of the previous Section.

We solved the CTRW model of transport in porous medium. This model was introduced on the basis of experimental observations and we believe that it can realistically capture lots of properties of the motion. We demonstrated that though the original model introduces dependence of velocity at consecutive steps of the walk (with finite probability velocity is conserved) we can reduce the model to the CTRW with independent steps by step's redefinition. Using exponential distribution of pore lengths observed in the experiment this redefinition of the step has the effect of an increase of the effective length of the step by a factor that characterizes the probability of velocity conservation at the junction. This increase factor has the empirically observed value of order $10$ so this point of our modelling is significant.

The decorrelation of velocity beyond a critical length which is much larger than the typical pore length $l_c$ seems to be true for a real porous medium. Indeed, consider the correlation function $\langle v(0)v(l)\rangle$ that describes correlation of velocities of the tracer particle at points separated by the distance $l$ passed along the trajectory. Here $v$ is certain component of the tracer velocity. This correlation function does not decay fast when $l\sim l_c$ because velocity at consequent pores is strongly correlated. However for large $l$ the correlation function decays. It is plausible that this decay is fast so the correlations can be neglected beyond finite correlation length. This length is then at least an order of magnitude larger than $l_c$ which is what we found in our model.

The reduced CTRW obtained in our model is separable: the length and velocity of the step are independent. Our solution assumes exponential distribution of step lengths and arbitrary distribution of the step velocities. We demonstrated in our derivation of the Montroll-Weiss equation that derivations can be performed for other than exponential distributions of the pore length. Study of the dependence of the anomalous diffusion's properties on the form of the PDF of the step's length is left for future work.

We considered a family of velocity statistics labeled by the parameter $\alpha$. That family derives from the assumption of Poiseuille profile in the channel \cite{Markus}. The parameter characterizes the probability of small velocities that determine near trapping of particles in the pore (the parameter gives also the stretched exponential tail of the velocity PDF). The resulting probability density function of the step duration has power-law tail with exponent smaller or equal $2$ so $\langle \tau\rangle=\infty$. The main reason for this power-law tail is the independence of length and velocity of the step which makes $\tau$ proportional to inverse velocity up to factor of length having but small fluctuations. It is this proportionality that underlies the anomalous scaling that we derived.

Inverse proportionality of $\tau$ and velocity implies that transport on long time-scales is determined by probabilities of small velocities. Qualitatively the flow at some junctions "does not know where to move" producing stagnation-type regions where the particle spends anomalously long times. These regions trap particles for long times forming anomaly in transport at long times, cf. with islands' formation of anomalous transport in Hamiltonian systems \cite{zaslavsky}.

Our solution demonstrates a clear difference between the cases where the PDF of velocity $p_v(v)$ has finite value at zero velocity, $\alpha<1$, and integrable power-law singularity, $1<\alpha<[1+\sqrt{5}]/2$, (the $\alpha=1$ is boundary case with logarithmic divergence). The average passed distance obeys
\begin{eqnarray}&& \!\!\!\!\!\!\!\!\!\!\!\!\langle l(t)\rangle\!\sim\!\frac{t\Gamma(1\!-\!\alpha\!+\!1/\alpha)}{\Gamma(1\!-\!\alpha)\ln t},\ \ \alpha<1;\ \ \langle l(t)\rangle\!\sim\!\frac{2t}{\ln^2 t},\ \ \alpha=1,\nonumber\\&&\!\!\!\!\!\!\!\!\!\!\!\!\left\langle l(t)\right\rangle\!\sim\!\frac{t^{1+\alpha-\alpha^2}\Gamma(1\!-\!\alpha\!+\!1/\alpha)}{\Gamma(\alpha\!-\!\alpha^2\!+\!2)\alpha|\Gamma(\alpha\!-\!\alpha^2)|\Gamma(\alpha^2\!-\!\alpha)},\label{finiti}\end{eqnarray} where the last formula holds at $1<\alpha\!<\![1+\sqrt{5}]/2$.

The results (\ref{finiti}) fit those obtained previously for $l-\tau$ CTRW. In the case where $\psi(\tau)\sim \tau^{-1-\beta}$ with $0<\beta<1$ it was found in \cite{Shl,Scher} that $\langle l(t)\rangle\propto t^{\beta}$. This was found for the mean position $\langle l\rangle$ of the propagating packet of carriers of current in amorphous semiconductors where the "mean flow" is an electric current \cite{Scher}. However, in our model at $1<\alpha\!<\![1+\sqrt{5}]/2$ the tail of $\psi(\tau)$ is proportional to $\tau^{-1-\beta}$ with $\beta=1+\alpha-\alpha^2$ so we have the same link between the decay exponent of $\psi(\tau)$ and growth exponent of $\langle l(t)\rangle$ in Eq.~(\ref{finiti}). Furthermore the $t/\ln t$ growth of $\langle l(t)\rangle$ is the same as that found in $l-\tau$ CTRW of \cite{Lester}.

We observe that for finite $p_v(v=0)$ the average passed distance obeys the expected (the average velocity is finite) linear growth in time up to logarithmic corrections. In contrast, in the case of diverging $p_v(v=0)$ the growth obeys a power-law with exponent smaller than one. This slowing of the growth is because of increase of trapping probability. At the upper limit $[1+\sqrt{5}]/2$ of physically admissible $\alpha$ the normalization of the PDF would diverge at zero velocity and $\langle l(t)\rangle$ would not grow at all describing trapped particle. (We observe that in a real medium some of the particles can stick to the boundaries producing $c\delta(v)$ term in $p_v(v)$ with $c<1$).

The difference between the cases of finite and infinite $p_v(v=0)$ continues to be strong when dispersion is considered. We find that,
\begin{eqnarray} && \!\!\!\!\!\!\!\!\!\!\!\!\!\!\! \frac{\langle l^2(t)\rangle-\langle l(t)\rangle^2}{\langle l(t)\rangle^2}\sim \frac{1}{\ln t}\left(1+\!\frac{4\psi_{\Gamma}(1\!-\!\!\alpha)}{\alpha}\right),\ \  \alpha<1,\nonumber\\&&\!\!\!\!\!\!\!\!\!\!\!\!
\frac{\langle l^2(t)\rangle-\langle l(t)\rangle^2}{\langle l(t)\rangle^2}\sim \frac{2}{\ln t},\ \  \alpha=1,\nonumber\\&&\!\!\!\!\!\!\!\!\!\!\!\!
\frac{\langle l^2(t)\rangle-\langle l(t)\rangle^2}{\langle l(t)\rangle^2}\sim\frac{2\Gamma^2(\alpha\!-\!\alpha^2\!+\!2)}{\Gamma(2\alpha-2\alpha^2+3)},\ \ \alpha>1.\end{eqnarray}
The law of large numbers implies that for $\alpha\leq 1$ the fluctuations of normalized distance $l(t)/\langle l(t)\rangle$ decay with time so that the limit in probability,
\begin{eqnarray}&&\!\!\!\!\!\!\!\!\!\!\!\!
\lim_{t\to\infty}\frac{l(t)}{\langle l(t)\rangle}=1,
\end{eqnarray}
holds. We find,
\begin{eqnarray}&&\
\lim_{t\to\infty}\frac{l(t)\Gamma(1-\alpha)\ln t}{\Gamma(1-\alpha+1/\alpha)t}=1,\ \ \alpha<1. \nonumber\\&&
\lim_{t\to\infty}\frac{l(t)\ln^2 t}{2t}=1, \ \ \alpha=1.
\end{eqnarray}
This can be considered as form of "ergodic theorem" that holds instead of finiteness of long-time limit of $l(t)/t$. The considered limits hold in probability, see the discussion of Eq.~(\ref{limitstrong}).

In contrast, in the $\alpha>1$ case $l(t)/\langle l(t)\rangle$ has order of one fluctuations. It seems plausible that in this case the PDF of $l(t)/\langle l(t)\rangle$ has finite $t\to\infty$ limit but the corresponding study is beyond our scope here. The result agrees with the one obtained in the $l-\tau$ CTRW where is was found that for $\psi(\tau)\sim \tau^{-1-\beta}$ tail with $0<\beta<1$ one has \cite{Shl,Scher},
\begin{eqnarray}&&\
\frac{\langle l^2(t)\rangle-\langle l(t)\rangle^2}{\langle l(t)\rangle^2}\sim\frac{2\Gamma^2(1+\beta)}{\Gamma(1+2\beta)}-1,
\end{eqnarray}
where in our case $\beta=1+\alpha-\alpha^2$ (the difference of $-1$ term from the result of \cite{Scher} seems to be a typo). Based on proved universality of the laws of growth of average and dispersion in $l-\tau$ CTRW is seems reasonable that $l-v$ CTRW also have universal behavior. Thus it seems plausible that anomalous diffusion exponents of the real porous medium could be predicted based on the tail of $\psi(\tau)$. The long tail is caused by stagnation points at junctions and/or the non-slip boundary conditions on the walls of the medium so that it could be universal (in the idealized model the tail would be $\tau^{-2}$, cf. \cite{Lester}). Thus, in experimental studies it becomes a crucial question to determine the tail of $\psi(\tau)$ and whether the tail's exponent is universal or several values are possible. In fact the results of \cite{Bijeljic} indicate that several values are possible.

It seems clear that the behavior of $p_v(v)$ at small $v$ will play significant role in future studies of transport in porous medium.

The described laws imply for the growth of dispersion in $\alpha<1$ case that the dispersion $\sigma_l^2(t)=\langle l^2(t)\rangle\!-\!\langle l(t)\rangle^2$ obeys,
\begin{eqnarray} && \!\!\!\!\!\!\!\!\!\!\!\!\!\!\! \!\sigma_l^2(t)\sim\! \frac{t^2}{\ln^3 t}\left(1\!+\!\frac{4\psi_{\Gamma}(1\!-\!\!\alpha)}{\alpha}\right)\frac{\Gamma^2(1\!-\!\alpha\!+\!1/\alpha)}{\Gamma^2(1\!-\!\alpha)}. 
\end{eqnarray}
This law of growth of dispersion proportionally to $t^2/\ln^3 t$ was observed in the $\tau-l$ CTRW model of tracer's motion in the porous medium considered in \cite{Lester}. In that model the distribution of $\tau$ was considered to have a $\tau^{-2}$ tail as in our case. The distribution of (vector) length of the step was considered to be fast decaying. These features of distributions of duration and length of the step hold in our case for $\alpha<1$ as well. The difference is that in our model $l$ and $v$ are independent rather than $l$ and $\tau$. Despite this difference the dispersion grows identically in the two models.

The dispersion in the $\alpha=1$ case reads,\begin{eqnarray} && \!\!\!\!\!\!\!\!\!\!\!\!\!\!\!
\langle l^2(t)\rangle-\langle l(t)\rangle^2\sim \frac{8t^2}{\ln^5 t},
\end{eqnarray}
and when $\alpha>1$,
\begin{eqnarray} && \!\!\!\!\!\!\!\!\!\!\!\!\!\!\!
\sigma_l^2(t)=\frac{2t^{2+2\alpha-2\alpha^2}\Gamma^2(1\!-\!\alpha\!+\!1/\alpha)}{\Gamma(2\alpha-2\alpha^2+3)
\alpha^2\Gamma^2(\alpha\!-\!\alpha^2)\Gamma^2(\alpha^2\!-\!\alpha)}.\end{eqnarray}
In the case of $\alpha=1$ the $\tau^{-2}$ tail has a logarithmic correction which introduces a logarithmic correction to the $t^2/\ln^3 t$ law that holds for the $\tau^{-2}$ tail. In the case $\alpha>1$ the power-law is slower than $\tau^{-2}$ and the dispersion grows as a power-law. It would be of interest to check if in the $\tau-l$ model of \cite{Lester} the use of $\alpha>1$ power-law for $\psi(\tau)$ would produce the same growth of dispersion.

We demonstrated that on long times scales the behaviors of the number of pores visited by the particle and of the passed distance coincide. This is because the pore length does not have strong fluctuations. This can be useful in further studies because number of visited pores is simpler for the study.

Our model is solved for an arbitrary distribution of velocities. It seems plausible that porous media occurring in nature can be modelled with power-law behavior at small velocities $p_v(v)\sim v^{-\delta}$ where $\delta$ has finite number of possible values depending on the formation history of the material. If the medium was formed by a (self-similar) fracturing process then we can expect a non-trivial delta different from zero. In contrast if the material was formed by gradual accumulation of randomly sized grains $p_v(v)\sim const$ could be describing the case (corresponding to $\tau^{-2}$ tail of the PDF of residence times). Once we know the form of $p_v(v)$ at small $v$ we can predict based on our solution the anomalous transport exponents. The experimental observation of $\delta$ is current work.

We disregarded the spatial geometry of the medium: only the magnitude of the passed distance was considered. This leaves outside of the study how much the tracer particle progressed along the major direction of the flow. The study of the corresponding correction factor (tortuosity) demands the introduction of vector length of the step or of the distributions of the corresponding projections of length and velocity. This study is left for future work.

We described three types of separable CTRWs of which two were used for the description of the tracer's motion in the porous medium: the $l-v$ model (for instance considered in this paper) and $l-\tau$ model (for instance considered in \cite{Lester}). It is a characteristic property of our model that the average residence time in the channel is infinite once $\langle v^{-1}\rangle$ diverges. This holds unless $p_v(v=0)=0$. Both cases of finite and vanishing $p_v(v=0)=0$ seem to be relevant in practice \cite{Bijeljic}. It seems though from preliminary observations that neither $l$ and $v$ or $\tau$ and $l$ can be considered independent. We thus think that the consideration of inseparable CTRW is needed for further progress. This will complicate the theory significantly necessitating the use of Fourier-Laplace transform of the propagator of inseparable walk provided in Section \ref{insep}. Further refinement of the use of CTRW for the description of tracer's motion in the porous medium is also needed for including the possibility of anamalously long "corridors" (or "preferential pathways") where the particle can pass anomalously long distances during one step of the walk. These corridors seem to be observed in experiments. This would demand introduction of the corresponding increased probability of large step length in $p(l)$. Our work thus presents a first step in the use of CTRW for the description of tracer's motion in the porous medium and provides meaningful directions for possible refinements extending the range of practical applications.

{\em Acknowledgements: } M.H. acknowledges support from the Swiss National Science Foundation (SNSF grant number 144645). I. F. thanks E. Barkai for helpful discussions of the CTRW.

\bibliography{Biblio}

\end{document}